\pdfoutput=1
\documentclass[12pt]{iopart}

\usepackage{graphicx}
\usepackage{abrev-jour-long}
\usepackage[square, sort&compress]{natbib}

\usepackage{color}

\begin{document}

\title{Self-gravitating razor-thin discs around black holes via multi-hole seeds}

\author{Ronaldo S. S. Vieira}

\address{Centro de Ci\^encias Naturais e Humanas, Universidade Federal do ABC, 09210-580 Santo Andr\'e, SP, Brazil}
\ead{ronaldo.vieira@ufabc.edu.br}
\vspace{10pt}
\begin{indented}
\item[]November 2019
\end{indented}

\begin{abstract}
We construct self-gravitating razor-thin discs of counterrotating matter around Schwarzschild black holes (BHs) by applying the ``displace, cut, and reflect'' method to known seed solutions representing multi-holes. All but one of the sources of the seed solution generate the surrounding annular disc, whereas the remaining BH is mapped onto a Schwarzschild BH which lies at the disc centre after the transformation. The discs are infinite in extent, have annular character, and are linearly stable up to the innermost stable circular orbit (ISCO) of the system. 
The spacetime is asymptotically flat, having finite Arnowitt-Deser-Misner (ADM) mass.
Moreover, all energy conditions for the disc are satisfied for radii larger than the ISCO radius; the method, however, produces counterrotating streams with superluminal velocities in the vicinity of the central BH.
We also comment on charged discs around extremal Reissner-Nordstr\"om BHs constructed from a Majumdar-Papapetrou $N$-BH seed solution. These simple examples can be extended to more general ``BH + disc'' solutions, obtained by the same method from seeds of the type ``BH + arbitrary axisymmetric source''. A natural follow-up of this work would be to construct discs around Reissner-Nordstr\"om BHs with arbitrary charge-to-mass ratio and around Kerr BHs. 
\end{abstract}

\submitto{Classical and Quantum Gravity}
\maketitle

%
%
%
%
%

\section{Introduction}

It is now widely accepted that black holes (BHs) are  ubiquitous in nature; moreover, recent gravitational-wave detections of binary BH mergers \cite{LIGO2016PhRvL, LIGO2016ApJL} and observations of the shadow of the central BH of M87 \cite{EHT2019paper1} confirmed with great precision that they behave as predicted by general relativity (see, however, \cite{bambiEtal2019PRD}). It is also a paradigm that BHs are generally surrounded by accretion discs. However, these discs are usually treated as test fluids \cite{abramowiczEtal1988ApJ, frank2002accretion, abramowiczFragileLRR, sadowskiEtal2015MNRAS, lasotaVieiraEtal2016AA}; it is clear that, in order to take into account the disc's self gravity, we must deal with distorted BH fields \cite{gerochHartle1982JMP}. Many efforts have been made in this direction, particularly in the quest for exact solutions of Einstein's field equations representing razor-thin discs \cite{morganMorgan1969PR, morganMorgan1970PRD, letelierOliveira1987JMP, bicakLyndenbellKatz1993PRD, gonzalezLetelier1999CQGra, gonzalezLetelier2000PRD, vogtLetelier2003PRD, ujevicLetelier2004PRD, gonzalezGutierrez2012CQGra,  vieiraLetelier2014GRG, gutierrezpineres2015GRG, gonzalezPimentel2016PRD, vieiraRamoscaroSaa2016PRD, semerak2016PRD, freitasSaa2017PRD} and  ``BH + razor-thin disc''  structures \cite{lemosletelier1993CQG, lemosLetelier1994PRD, lemosLetelier1996IJMPD, saaVenegeroles1999PhLA,  semerakZacek2000CQGra, semerakZacek2000PASJ, semerak2002review, zacekSemerak2002CzJP,  karasHureSemerak2004CQGra, vogtLetelier2005PRD, gutierrezpineresEtal2014IJMPD} 
\footnote{One of the first axially symmetric solutions of Einstein's equations was obtained by Curzon \cite{curzonPLMS1924, griffithsPodolsky2009exact}, whose rotating extension was obtained in \cite{halilsoyJMP1992}.}.
Charged discs around extremal Reissner-Nordstr\"om \mbox{(R-N)} BHs are also recently getting attention \cite{loraclavijo-ospinahenao-pedraza2010PRD, semerak2016PRD,  polcarSukovaSemerak2019ApJ}. 
Also, the dynamics of test particles in these systems was widely studied \cite{saaVenegeroles1999PhLA, semerakZacekZellerin1999MNRAS, semerakSukova2010MNRAS, semerakSukova2012MNRAS, sukovaSemerak2013MNRAS,  witzanySemerakSukova2015MNRAS, vieiraRamoscaroSaa2016PRD, polcarSukovaSemerak2019ApJ}. 
Therefore, in addition to their important theoretical interest, exact ``BH + disc'' solutions of Einstein's field equations became a crucial tool to understand the effects of the self gravity of accretion discs around astrophysical BHs.

We present in this work a new family of exact solutions of Einstein's field equations representing annular razor-thin discs around Schwarzschild BHs. The method presented here provides a simple procedure to construct these composite systems once we have an adequate seed solution representing a Schwarzschild BH plus an arbitrary external axisymmetric source. It also allows us to construct charged discs around extremal R-N BHs.  

Section~\ref{sec:RTDs} gives a brief overview of the formalism used to construct the razor-thin disc structures. Section~\ref{sec:AnnularSchwarzschild} presents the central result of the paper: the construction of annular razor-thin discs around Schwarzschild BHs via multi-hole seed solutions. Section~\ref{sec:Charged} briefly comments on the  extreme R-N case. We present our conclusions in Section~\ref{sec:Discussion}.

\section{Razor-thin discs from seed solutions}
\label{sec:RTDs}

Let us assume we have a spacetime composed of a BH plus an external axisymmetric source, represented by the Weyl metric 
\begin{equation}\label{eq:Weylmetric}
ds^2=-e^{2\psi}\,dt^2 + e^{-2\psi}\,[\,e^{2\gamma}(\,d\rho^2 + dz^2\,) + \rho^2 d\varphi^2 \,]\,,
\end{equation}
which we will denominate a ``seed'' solution. Interpreting $z$ as a ``vertical'' coordinate, we may apply a Kuzmin-like transformation \cite{binneytremaineGD} to the metric functions of the seed solution in order to obtain a razor-thin disc source, meaning a delta-like distributional source on the symmetry plane of the new system. This procedure is often called the ``displace, cut, and reflect'' (DCR) method \cite{vogtLetelier2003PRD, vieiraLetelier2014GRG, gonzalezPimentel2016PRD, freitasSaa2017PRD, navarronogueraEtal2018GRG} and consists of making the transformation
\begin{equation}\label{eq:DCRtransformation}
z\to |z|+\lambda
\end{equation}
to the metric functions $\psi$ and $\gamma$, with $\lambda>0$. 
As we will see, we can construct self-gravitating discs around a central Schwarzschild BH by a suitable choice of the seed solution and of the DCR parameter $\lambda$.

\begin{figure*}
\begin{center}
\includegraphics[width=\textwidth]{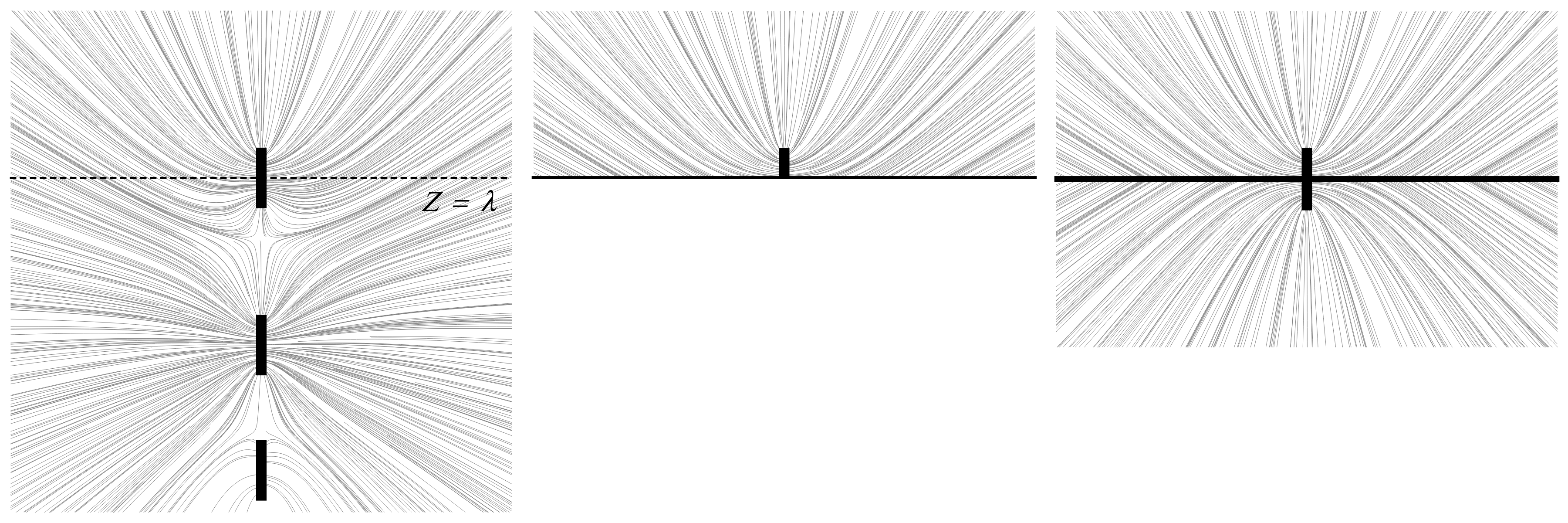}
\caption{The DCR method applied to a 3-rod seed solution ($N=3$, see Section \ref{sec:AnnularSchwarzschild}); we choose $L_1=2m_1$ in order to have a Schwarzschild BH as the highest seed source (see text). 
The system's field lines, defined in Section \ref{sec:selfgravity}, are drawn in gray; we note that the field lines are perpendicular to the contour lines of the metric function $\psi$. The first step is to ``displace'' the $z$-coordinate by an amount $\lambda=Z_1$ (left panel). We then ``cut'' the spacetime at $z=\lambda=Z_1$  in such a way that the highest Schwarzschild rod is cut in half. We keep the upper part of the spacetime (middle panel) and ``reflect'' it with respect to the horizontal black line (right panel). The result is a Schwarzschild BH of mass $m_1$ (represented by the vertical rod) surrounded by an annular razor-thin disc (represented by the horizontal black line). The procedure is equivalent to applying the DCR transformation (\ref{eq:DCRtransformation}) to the seed metric functions of Section \ref{sec:AnnularSchwarzschild}, with $\lambda=Z_1$.
}
\label{fig:DCRmethod}
\end{center}
\end{figure*}

The necessary formalism to deal with relativistic razor-thin discs is the theory of distributional sources with support on timelike spacetime hypersurfaces \cite{taub1980JMP, lemosLetelier1994PRD, vieiraLetelier2014GRG, vieiraRamoscaroSaa2016PRD, freitasSaa2017PRD}. The stress-energy tensor is written as $T^\mu_{\ \nu} = Q^\mu_{\ \nu}\,\hat\delta(z) + D^\mu_{\ \nu}$,
where $ \hat\delta(z) = \delta(z)/\sqrt{g_{zz}}$ is the covariant delta distribution in curved spacetimes \cite{vieiraLetelier2014GRG, vieiraRamoscaroSaa2016PRD}, $Q^\mu_{\ \nu}$ is the disc's stress-energy tensor, and $D^\mu_{\ \nu}$ represents the smooth matter-energy content of the spacetime (such as a halo or a thickened disc, for instance). The form of the metric allows us to write $Q^\mu_{\ \nu} = diag(-\sigma, P_\rho, P_\varphi, 0)$, the proper surface energy density $\sigma$ and the principal pressures $P_\rho$ (radial) and $P_\varphi$ (azimuthal) of the disc.
It is worth noting that, when we consider the covariant delta distribution $\hat\delta$, the formalism gives us, in a self-consistent manner, the ``physical'' (or ``true'') stress-energy tensor $Q^\mu_{\ \nu}$ of the disc \cite{vieiraLetelier2014GRG, vieiraRamoscaroSaa2016PRD} without the need to differentiate between the ``formal'' and the ``true'' stress-energy tensors mentioned for instance in \cite{lemosLetelier1994PRD, gonzalezLetelier1999CQGra, gonzalezLetelier2000PRD, vogtLetelier2003PRD, gutierrezpineresEtal2014IJMPD, vieiraLetelier2014GRG, gonzalezPimentel2016PRD, navarronogueraEtal2018GRG}.

In terms of the metric derivatives, the surface density $\sigma$ and azimuthal pressure $P_\varphi$ of the disc are written as \cite{vieiraRamoscaroSaa2016PRD} 
\begin{equation}\label{eq:sigmaWeyl}
\sigma= \frac{e^{\psi-\gamma}}{4\pi}\left(2\frac{\partial\psi}{\partial |z|} - \frac{\partial\gamma}{\partial |z|}\right)\Bigg|_{z=0}\,,
\end{equation}
\begin{equation}\label{eq:PphiWeyl}
P_\varphi= \frac{e^{\psi-\gamma}}{4\pi}\frac{\partial\gamma}{\partial |z|}\Bigg|_{z=0}\,.
\end{equation}
Static razor-thin discs described by a Weyl metric have radial pressure $P_\rho=0$, and therefore are made of counterrotating streams \cite{morganMorgan1969PR, lemosLetelier1994PRD, klein1997CQG, gonzalezEspitia2003PRD, freitasSaa2017PRD}. Moreover, the DCR procedure keeps unaltered the ``upper'' part of the seed spacetime, above the disc; in particular, the stress-energy tensor of the upper part is preserved, as well as any singularities/black holes. These structures are then reflected to the ``lower'' part of the new spacetime, below the disc. The procedure is illustrated in Figure~\ref{fig:DCRmethod} for the $N$-rod seed solution \cite{letelierOliveira1998CQG} mentioned in Section \ref{sec:AnnularSchwarzschild}.


\section{Annular discs around Schwarzschild BHs}
\label{sec:AnnularSchwarzschild}

We consider as seed the ``$N$ collinear Schwarzschild BHs'' solution \cite{israelKhan1964NCim} or, more generally, the ``$N$ collinear Weyl rods'' solution \cite{letelierOliveira1998CQG}. 
The metric has the Weyl form (\ref{eq:Weylmetric}) with functions $\psi$ and $\gamma$ given by \cite{letelierOliveira1998CQG}
\begin{equation}\label{eq:psiSeed}
\psi(\rho,z)=\sum_{i=1}^N \psi_i(\rho,z)\,,
\end{equation}
where
\begin{equation}\label{eq:psi-i-Seed}
\psi_i(\rho,z)=\frac{m_i}{L_i}{\rm log}\left[\frac{R_{i}^+ + R_{i}^- - L_i}{R_{i}^+ + R_{i}^- + L_i} \right]\,,
\end{equation}
and
\begin{equation}\label{eq:gammaSeed}
\gamma(\rho,z)=\sum_{i=1}^N\sum_{j=1}^N
\frac{m_i\, m_j}{L_i\, L_j}
{\rm log}\left[\frac{E_{i\,j}^{+-}\,E_{i\,j}^{-+}}{E_{i\,j}^{++}\,E_{i\,j}^{--}} \right]\,,
\end{equation}
with
\begin{equation}\label{eq:Eij}
E_{i\,j}^{\pm\pm}=R_i^\pm R_j^\pm + z_i^\pm z_j^\pm + \rho^2\,.
\end{equation}
Here, 
\begin{equation}\label{eq:Ri}
R_{i}^\pm=\sqrt{\rho^2 + (z_i^\pm)^2}
\end{equation}
and
\begin{equation}\label{eq:zi}
z_{i}^\pm=z - \Big(Z_i \mp \frac{1}{2}L_i\Big)\,.
\end{equation}
The function $\psi$ represents the gravitational potential of $N$ one-dimensional rods of length $L_i$ and mass $m_i$ along the $z$-axis, each one with centre at at $z=Z_i$ \cite{letelierOliveira1998CQG}.
We order the $Z_i$ such that $Z_i-L_i/2>Z_j+L_j/2$ if $i<j$, meaning that there will be no intersection between the rods. They will have $Z_1>Z_2>...>Z_N$, in such a way that the rod labeled by the index 1 will be at highest $z$ among all sources. When $L_k=2m_k$, the $k$-th Weyl rod reduces to a Schwarzschild BH of mass $m_k$ \cite{israelKhan1964NCim, letelierOliveira1998CQG}.

The case $\lambda>Z_1+L_1/2$ generates a disc-only solution and was treated in \cite{bicakLyndenbellKatz1993PRD}. The case $Z_2+L_2/2<\lambda<Z_1-L_1/2$ generates two equal-mass BHs with a disc in their mid-plane; however, the DCR method will, in general, preserve a ``strut'' \cite{israelKhan1964NCim, synge1960relativity} connecting the two BHs (and therefore crossing the disc). We regard this solution as unphysical.

In order to construct annular discs around Schwarzschild BHs, let us consider the $N$-rod seed solution presented above, with $L_1=2 m_1$ in such a way that the first rod is a Schwarzschild BH with mass $m_1$ (we assume $Z_1>0$). If we choose the DCR parameter as $\lambda=Z_1$, the resulting spacetime will be composed of a Schwarzschild BH of mass $m_1$ surrounded by a razor-thin disc generated by the other ($N-1$) sources of the seed solution, as shown below.

Intuitively, the DCR method will preserve the upper part of the rod, source of the seed potential $\psi_1$, and reflect it with respect to the horizontal plane, generating in the DCR-transformed spacetime an equal rod but now encircled by a razor-thin disc (see Figure~\ref{fig:DCRmethod}). This rod will then be a Schwarzschild BH of masss $m_1$. Indeed, putting $\lambda=Z_1$, the term $\psi_1$ will be a function of the sum $R_{1}^+ + R_{1}^-$, where
\begin{equation}
R_{1}^\pm=\sqrt{\rho^2 + (z_1^\pm)^2}
\end{equation}
and
\begin{equation}
z_{1}^\pm= |z| \mp \frac{1}{2}L_1
\end{equation}
after the DCR transformation. If $z\geq 0$ we have the usual Schwarzschild field. Now, if $z<0$, we have $z_{1}^\pm= -z \mp \frac{1}{2}L_1$, implying $(z_1^\pm(-z))^2=(z_1^\mp(z))^2$ and thus $R_{1}^\pm(\rho,-z)=R_{1}^\mp(\rho,z)$. Therefore $R_{1}^+(\rho,-z) + R_{1}^-(\rho,-z) = R_{1}^-(\rho,z) + R_{1}^+(\rho,z)$ and $\psi_1(\rho,-z)=\psi_1(\rho,z)$. It follows that $\psi_1$ 
is correctly identified with the field of a Schwarzschild BH of mass $m_1$.
Moreover, the other functions $\psi_i$, $i>1$, will contain linear terms in $|z|$ when expanded around $z=0$. They give, in this way, a direct contribution to the disc's stress-energy tensor, according to (\ref{eq:sigmaWeyl})--(\ref{eq:PphiWeyl}).

Although the explicit expressions for the surface density and azimuthal pressure of the disc are too lengthy to be presented here, they can be obtained straightforwardly from the DCR procedure~(\ref{eq:DCRtransformation}) with $\lambda = Z_1$ applied to the seed metric functions $\psi$ and $\gamma$ (\ref{eq:psiSeed})--(\ref{eq:zi}), by means of expressions (\ref{eq:sigmaWeyl})--(\ref{eq:PphiWeyl}).
The resulting metric functions are given by Equations (\ref{eq:psiSeed})--(\ref{eq:Ri})  but now with 
\begin{equation}
z_i^\pm = |z| + Z_1 - \Big(Z_i \mp \frac{1}{2}L_i\Big)\,.
\end{equation}
The disc's surface density $\sigma$ always vanishes at the event horizon ($\sigma(0)=0$ in Weyl coordinates) and the azimuthal pressure $P_\varphi$ satisfies, in general, $P_\varphi(0)>0$.
The discs have infinite extent;
their density profile goes asymptotically as $\sigma\sim1/\rho^3$ (as in the case of the classical Kuzmin disc) and their pressure profile as $P_\varphi\sim1/\rho^4$. Accordingly, the circular speed of the counterrotating streams ($V^2=P_\varphi/\sigma$, see \cite{bicakLyndenbellKatz1993PRD, lemosLetelier1994PRD, gonzalezEspitia2003PRD}) has a Keplerian asymptotic profile $V\sim\rho^{-1/2}$. We always have $V^2<1$ outside the marginally stable orbit (see below), in such a way that the streams are physically plausible (although we may have $V^2>1$ in regions very close to the BH, see for instance Figure~\ref{fig:figN2a}; these regions, however, are not considered as part of the disc, as explained below).
We also note that the property $\sigma(0)=0$ of the ``BH + disc'' solution is due to relativistic effects, being related to the existence of an event horizon; the corresponding superposition ``monopole + $N$ Kuzmin potentials'' in Newtonian gravity always gives us a positive surface density at the disc centre (if all mass parameters are positive).

Moreover, the structure of the metric functions guarantees us that the ``BH + disc'' spacetime is asymptotically flat. Therefore its Arnowitt-Deser-Misner (ADM) mass \cite{poisson2004toolkit} is finite. It is given by $M_{\rm ADM}=\sum_{i=1}^N m_i$ and coincides with the ADM mass of the seed spacetime. Other definitions for the disc mass \cite{vogtLetelier2003PRD, vieiraLetelier2014GRG} give us a finite total mass for the disc itself. 
We restrict ourselves to models with $M_{\rm ADM}>m_1$; this condition can be interpreted as a ``positive total mass'' for the disc.


\subsection{Stability of circular geodesics}
 
Vieira et al. \cite{vieiraRamoscaroSaa2016PRD} found a sufficient condition for the vertical stability of timelike circular geodesics on the equatorial plane of the system, if it contains a razor-thin disc. They provided a definite vertical stability criterion for these orbits, extension of the Newtonian criterion developed in \cite{vieiraRamosCaro2016CeMDA}. It was obtained that if the strong energy condition for the disc fluid is satisfied, timelike circular geodesics are always stable under small vertical perturbations (in the $z-$ direction). Since in our case the strong energy condition for the disc is given by $\sigma + P_\varphi>0$ and this quantity is always positive for the discs considered here (see below), all timelike circular geodesics are vertically stable. 

On the other hand, Rayleigh's radial stability criterion \cite{letelier2003PRD, abramowiczKluzniak2005ApSS, vieiraMarekEtal2014PRD} (which states that if the angular momentum $L_z(\rho)$ of timelike circular geodesics satisfies $(L^2_z)_{,\rho}>0$ then these orbits are stable under small radial perturbations) shows us that the disc has an innermost stable circular (geodesic) orbit (ISCO) in the presence of the BH. In this ``particle'' approximation, consistent with the interpretation of the azimuthal pressure $P_\varphi$ as being due to counterrotating geodesic streams with the same speed $V^2=P_\varphi/\sigma$ \cite{gonzalezEspitia2003PRD}, the disc may be regarded as linearly stable for radii larger than the ISCO radius $\rho_{ISCO}$.
We argue below that, by an adequate choice of parameters, the disc density, azimuthal pressure, and its influence on the system's field lines are negligible for radii smaller than $\rho_{ISCO}$.
We note that the discs described below satisfy all energy conditions for $\rho\geq\rho_{ISCO}$.

\subsection{The $N=2$ and $N=3$ discs}

\begin{figure*}
\begin{center}
\includegraphics[width=0.48\columnwidth]{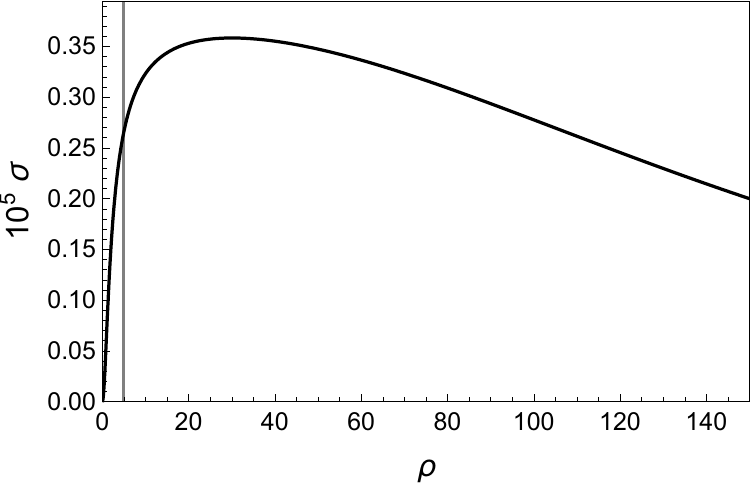}\quad
\includegraphics[width=0.48\columnwidth]{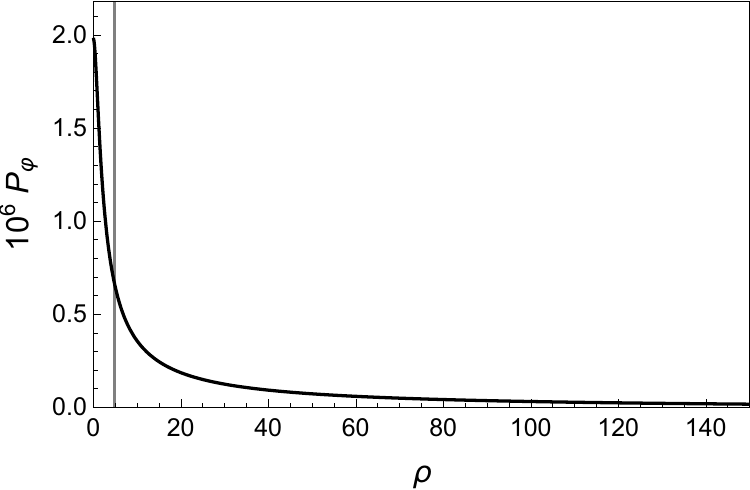}
\\ \hfill\\
\includegraphics[width=0.48\columnwidth]{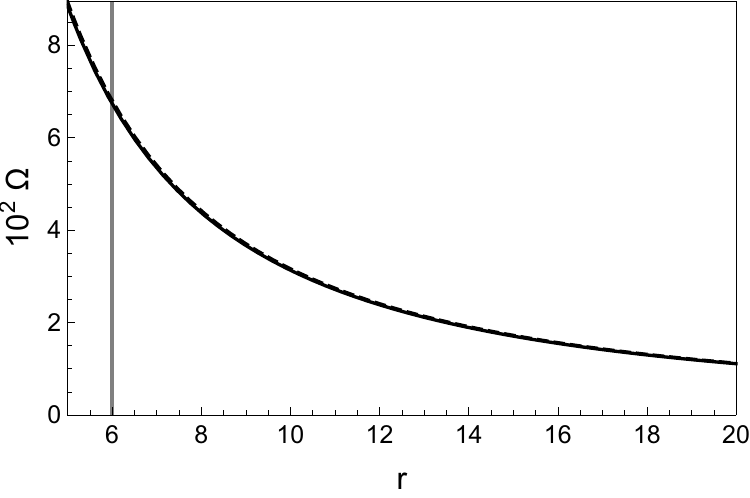}\quad
\includegraphics[width=0.48\columnwidth]{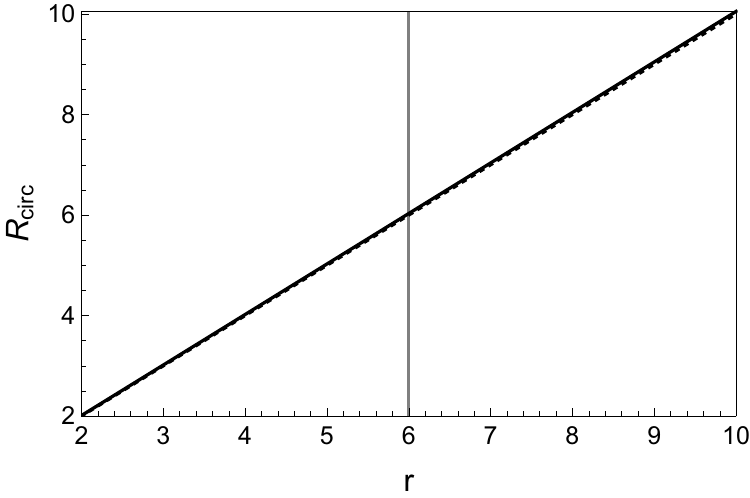}
\\ \hfill\\
\includegraphics[width=0.48\columnwidth]{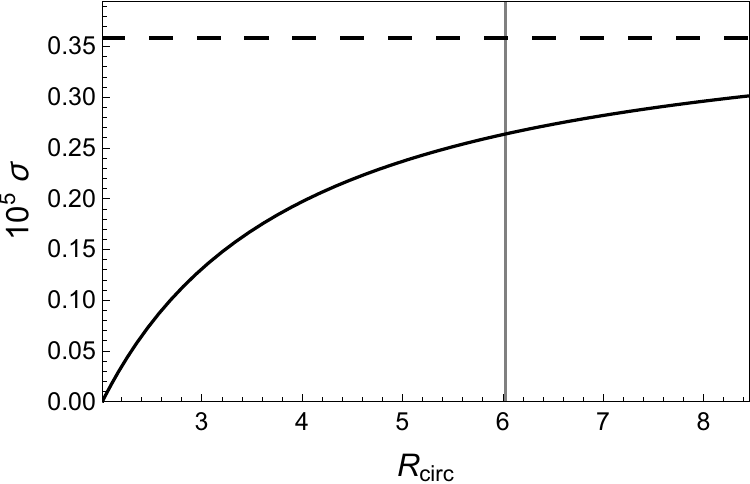}\quad
\includegraphics[width=0.48\columnwidth]{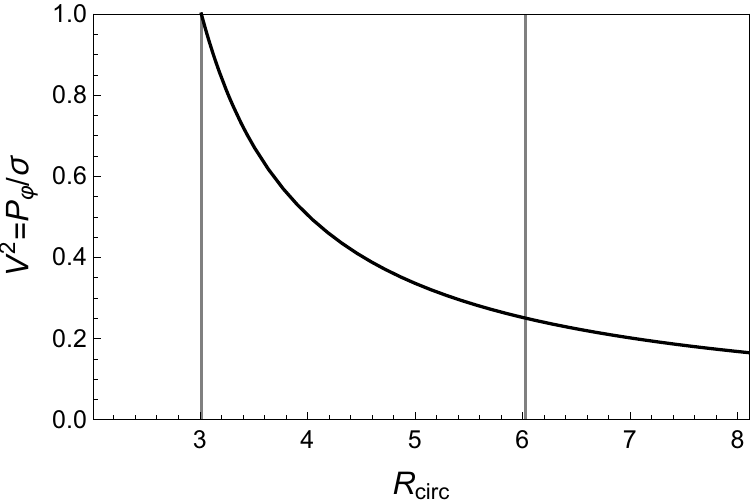}
\caption{
Top left: 
surface energy density of the $N=2$ ``Schwarzschild BH + disc'' system, with parameters $m_1=1$, $m_2=1$, $L_1=2$, $L_2=2$, $Z_1=\lambda=200$, $Z_2=0$. Top right: corresponding azimuthal pressure. The vertical line in the top panels represents $\rho_{ISCO}$. All energy conditions are satisfied for $\rho>\rho_{ISCO}$. 
Middle left: angular velocity profile for the disc (solid curve) as a function of Schwarzschild-like radius $r$ (defined and discussed in Section~\ref{sec:selfgravity}), and corresponding Keplerian profile (dashed curve). Middle right: circumferential radius $R_{\rm circ}$ (defined and discussed in Section~\ref{sec:nearHorizon}) in the presence of the disc as a function of $r$ (solid curve) and the corresponding function for Schwarzschild spacetime (dashed curve). The vertical line represents $r_{ISCO}$ in both panels.
Bottom left: surface density as a function of $R_{\rm circ}$ in the region near the horizon. The dashed line represents the maximum surface density $\sigma_{\rm max}$.  We have $\sigma_{ISCO}/\sigma_{\rm max} \approx 0.736$. Bottom right: circular velocity profile $V^2 = P_\varphi/\sigma$  as a function of $R_{\rm circ}$. The vertical line near $R_{\rm circ} = 6$ represents $R_{ISCO}$, while the vertical line near $R_{\rm circ} = 3$ represents the $V^2 = 1$ circumferential radius.
}
\label{fig:figN2a}
\end{center}
\end{figure*}

\begin{figure*}
\begin{center}
\includegraphics[width=0.48\columnwidth]{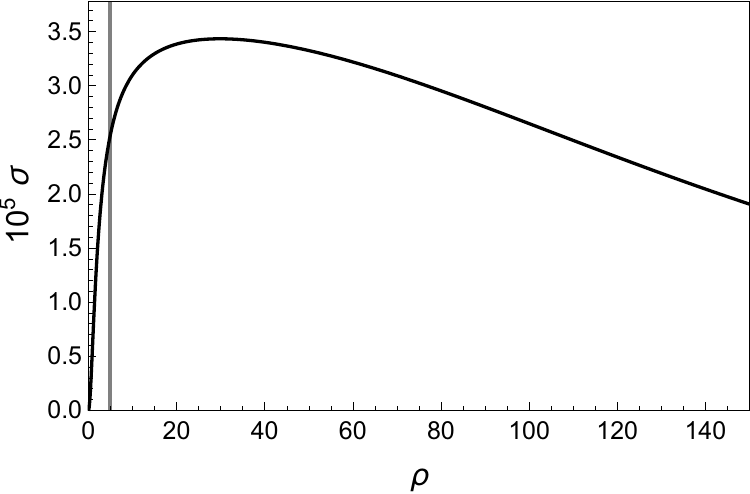}\quad
\includegraphics[width=0.48\columnwidth]{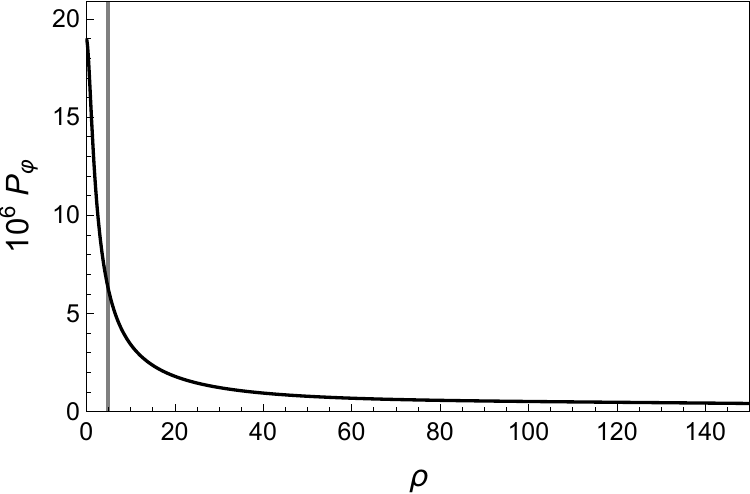}
\\ \hfill\\
\includegraphics[width=0.48\columnwidth]{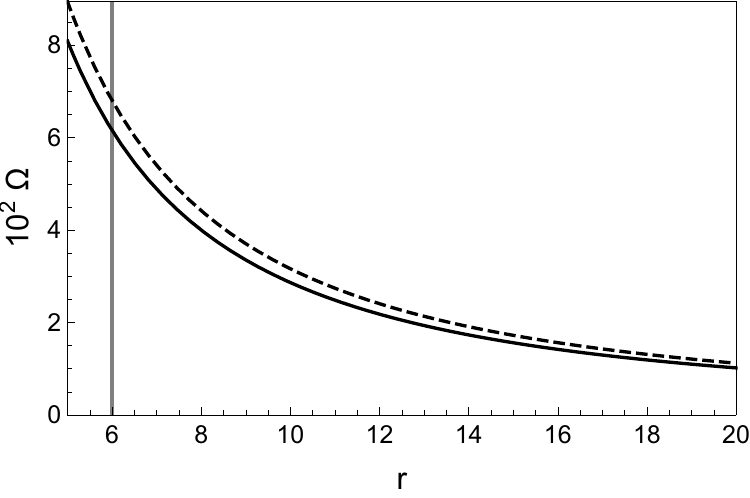}\quad
\includegraphics[width=0.48\columnwidth]{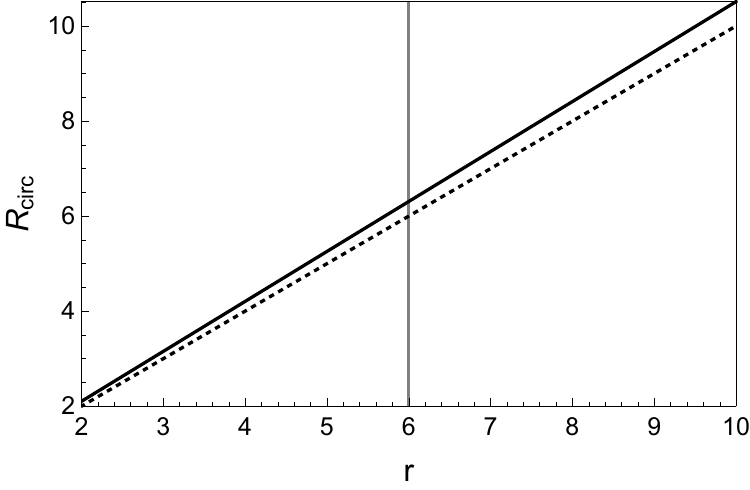}
\\ \hfill\\
\includegraphics[width=0.48\columnwidth]{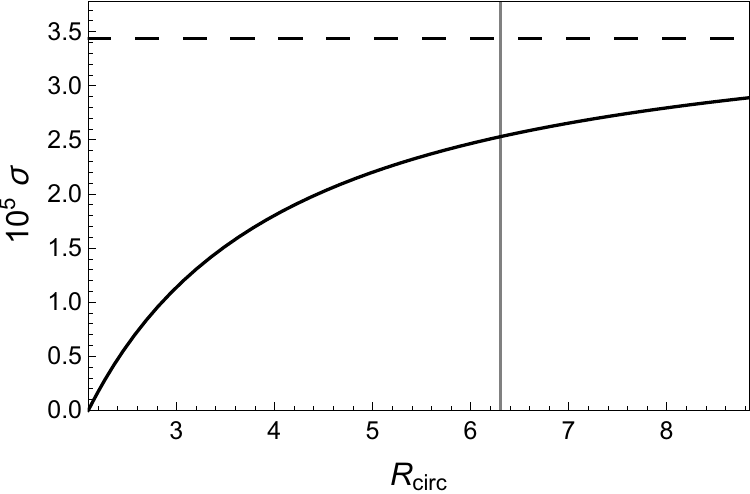}\quad
\includegraphics[width=0.48\columnwidth]{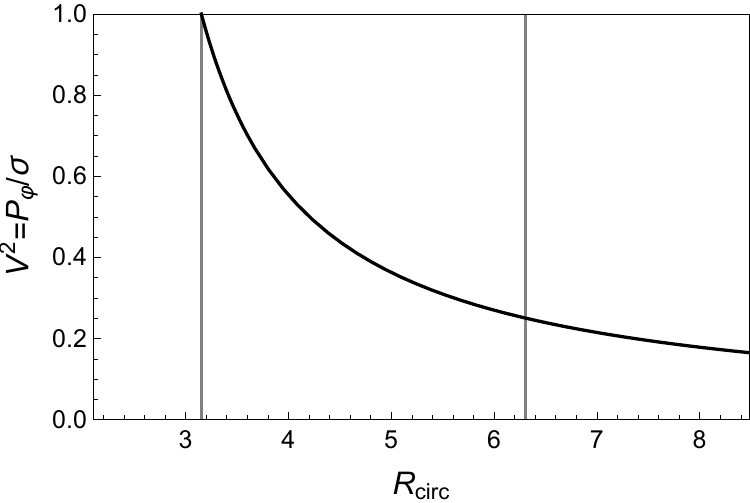}
\caption{
Same as Figure~\ref{fig:figN2a}, but for a $N=2$ ``Schwarzschild BH + disc'' system with parameters $m_1=1$, $m_2=10$, $L_1=2$, $L_2=20$, $Z_1=\lambda=200$, $Z_2=0$. We have $\sigma_{ISCO}/\sigma_{\rm max} \approx 0.736$.
}
\label{fig:figN3a}
\end{center}
\end{figure*}

\begin{figure*}
\begin{center}
\includegraphics[width=0.48\columnwidth]{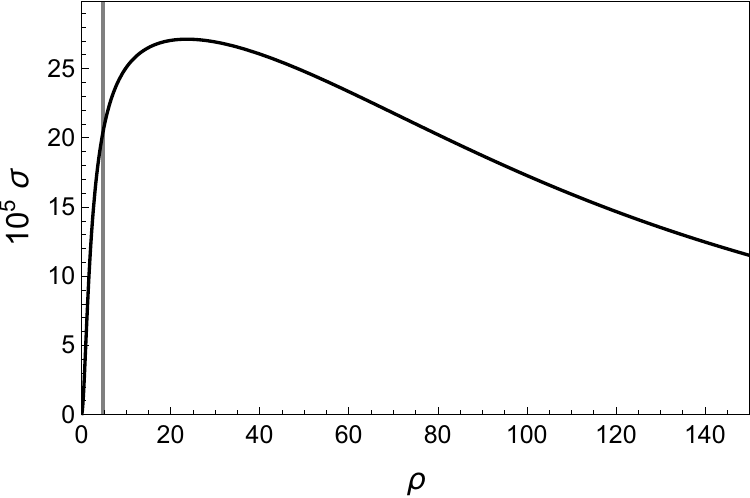}\quad
\includegraphics[width=0.48\columnwidth]{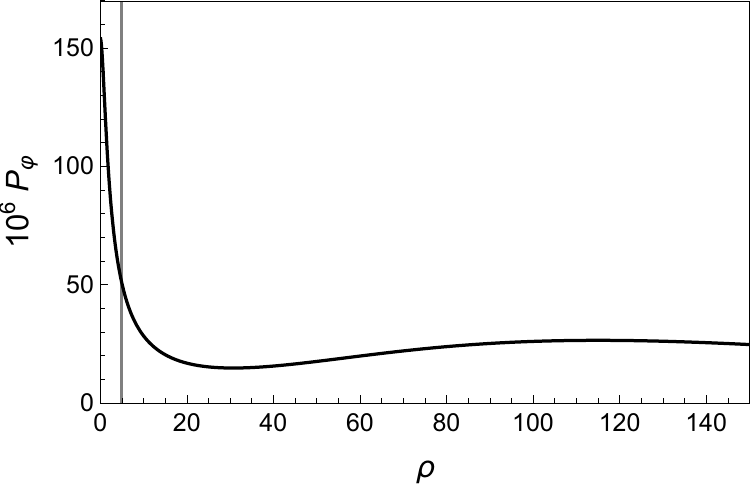}
\\ \hfill\\
\includegraphics[width=0.48\columnwidth]{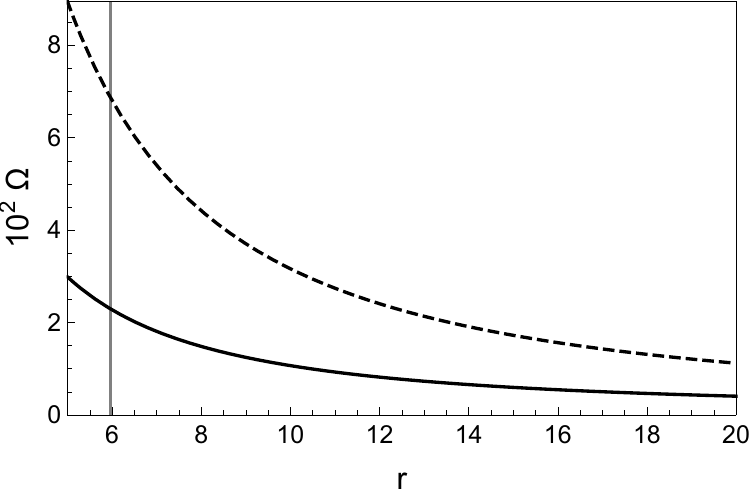}\quad
\includegraphics[width=0.48\columnwidth]{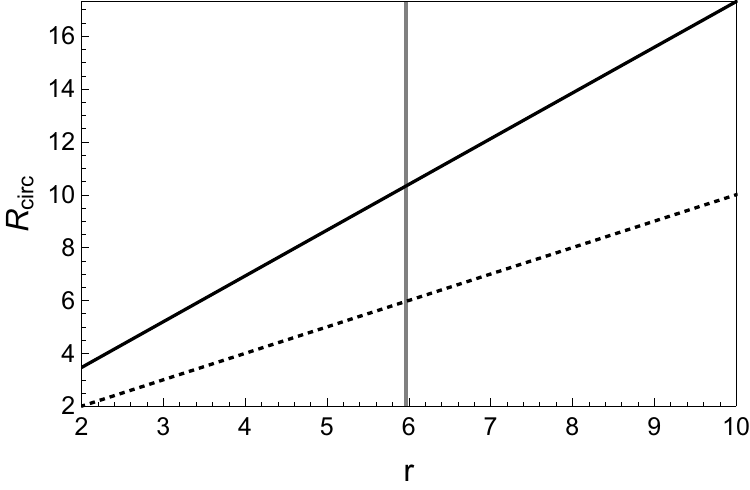}
\\ \hfill\\
\includegraphics[width=0.48\columnwidth]{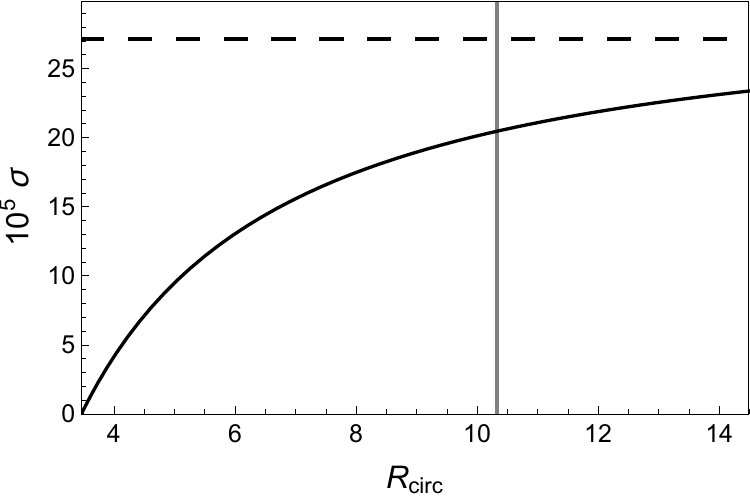}\quad
\includegraphics[width=0.48\columnwidth]{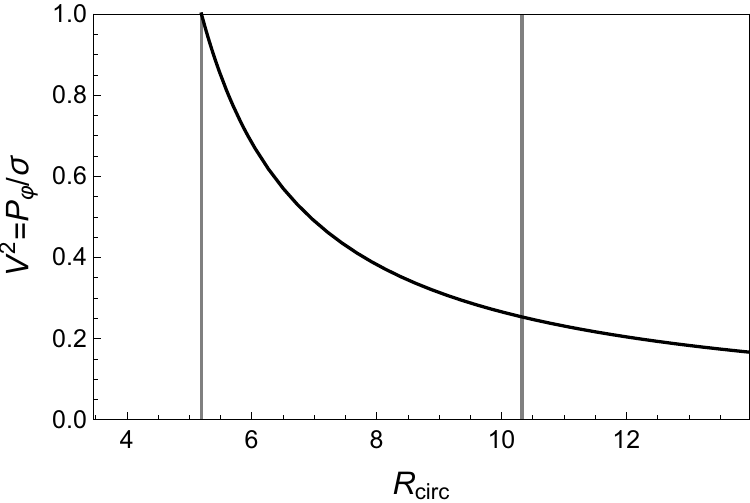}
\caption{
Same as Figure~\ref{fig:figN2a}, but for a $N=2$ ``Schwarzschild BH + disc'' system with parameters $m_1=1$, $m_2=100$, $L_1=2$, $L_2=200$, $Z_1=\lambda=200$, $Z_2=0$. We have $\sigma_{ISCO}/\sigma_{\rm max} \approx 0.754$.
}
\label{fig:figN4a}
\end{center}
\end{figure*}

The simplest solution is the $N=2$ case, with $L_2,\, m_2>0$, in which the disc is generated by only one seed source $m_2$. The general picture in this case is of a disc around a BH of mass $m_1$ with a regular density profile satisfying $\sigma(0)=0$ and peaked at $\rho>0$ (we fix $m_1 = 1$ and $L_1 = 2m_1$ throughout the paper, without loss of generality). The pressure profile always has $P_\varphi(0)>0$ and $P_\varphi$ is at least one or two orders of magnitude smaller than $\sigma$. 
We present the density and pressure profiles of the discs, for mass parameters $m_2 = 1$, $m_2 = 10$, and $m_2 = 100$, in Figures~\ref{fig:figN2a}, \ref{fig:figN3a}, and \ref{fig:figN4a}, respectively (with $L_2 = 2 m_2$). We adopt $Z_2 = 0$ and $\lambda =Z_1 = 200$, so that $\lambda$ is the vertical coordinate distance between the two seed BHs.

\begin{figure*}
\begin{center}
\includegraphics[width=0.48\columnwidth]{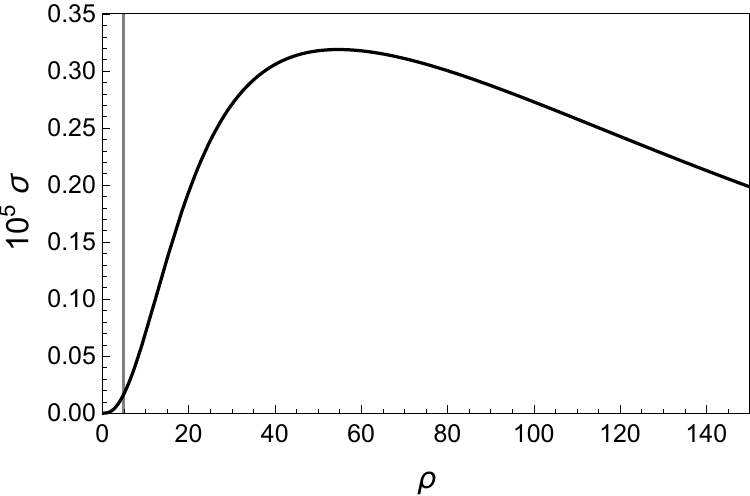}\quad
\includegraphics[width=0.48\columnwidth]{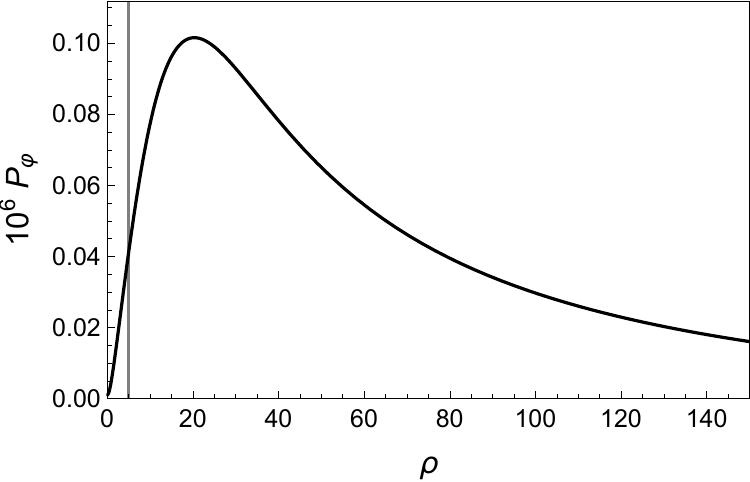}
\\ \hfill\\
\includegraphics[width=0.48\columnwidth]{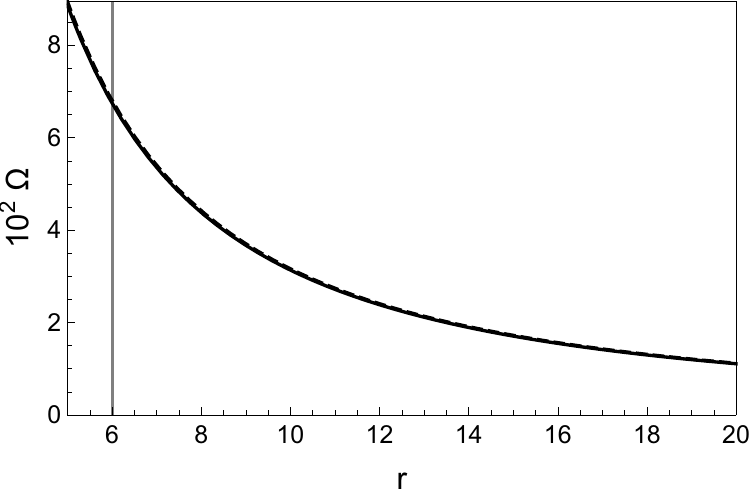}\quad
\includegraphics[width=0.48\columnwidth]{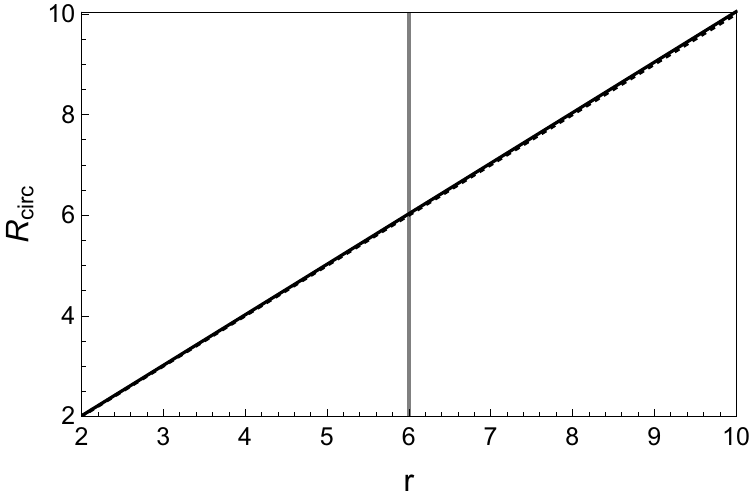}
\\ \hfill\\
\includegraphics[width=0.48\columnwidth]{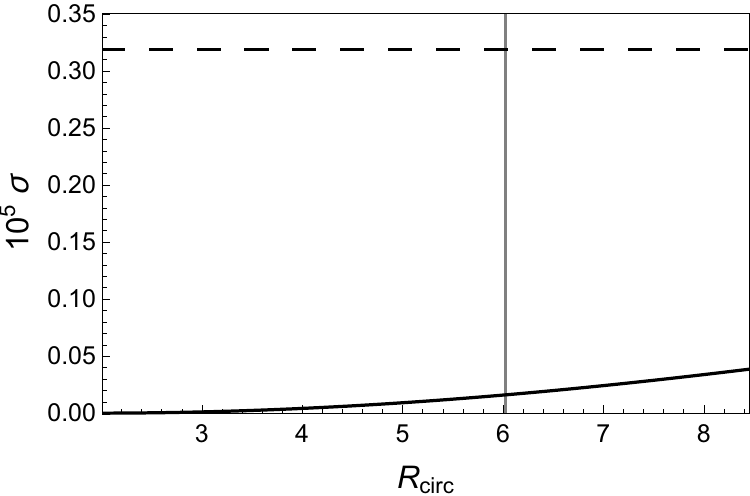}\quad
\includegraphics[width=0.48\columnwidth]{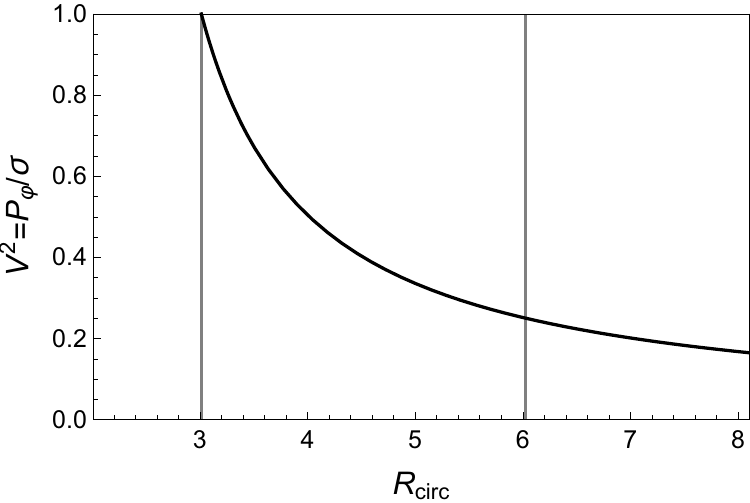}
\caption{
Same as Figure~\ref{fig:figN2a}, but for a $N=3$ ``Schwarzschild BH + disc'' system with parameters $m_1=1$, $m_2=-0.01397$, $m_3=1$, $L_1=2$, $L_2=1$, $L_3=2$, $Z_1=\lambda=200$, $Z_2=176.35$, $Z_3=0$. We have $\sigma_{ISCO}/\sigma_{\rm max} \approx 0.05$. Compare with Figure~\ref{fig:figN2a}, whose configuration is the same but without the negative-mass rod.
}
\label{fig:figN5a}
\end{center}
\end{figure*}

\begin{figure*}
\begin{center}
\includegraphics[width=0.48\columnwidth]{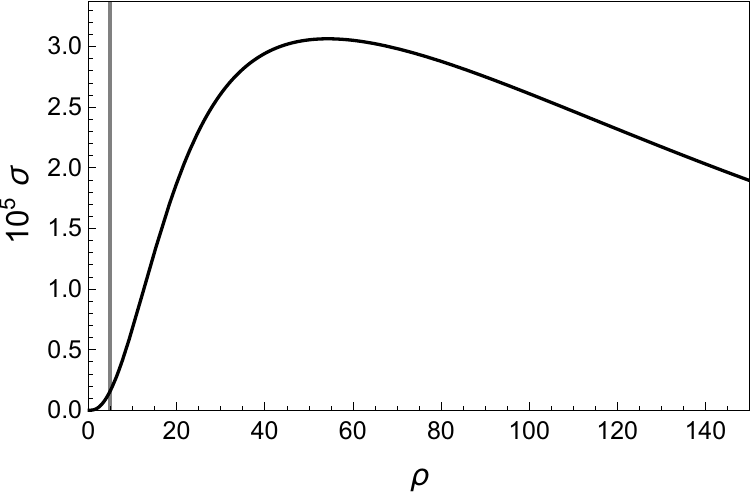}\quad
\includegraphics[width=0.48\columnwidth]{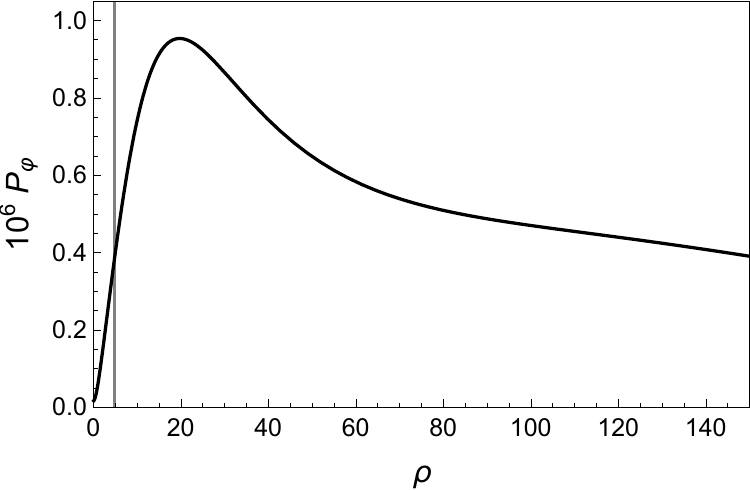}
\\ \hfill\\
\includegraphics[width=0.48\columnwidth]{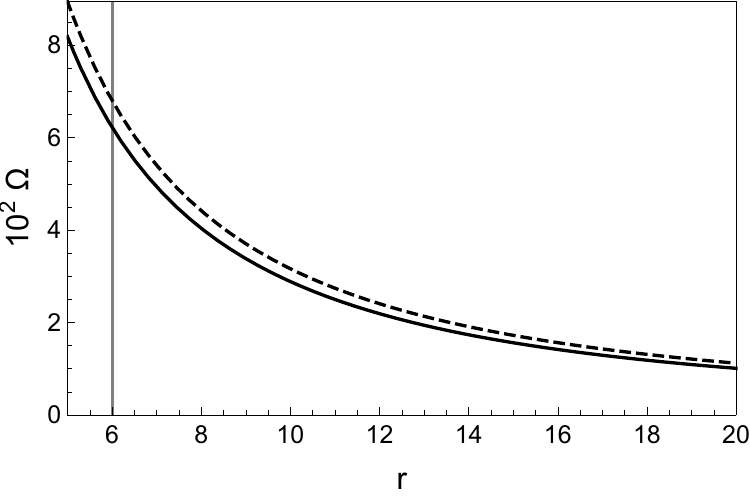}\quad
\includegraphics[width=0.48\columnwidth]{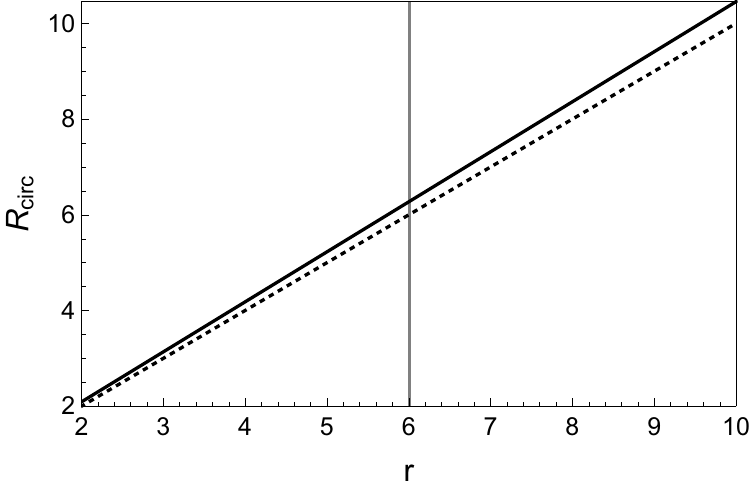}
\\ \hfill\\
\includegraphics[width=0.48\columnwidth]{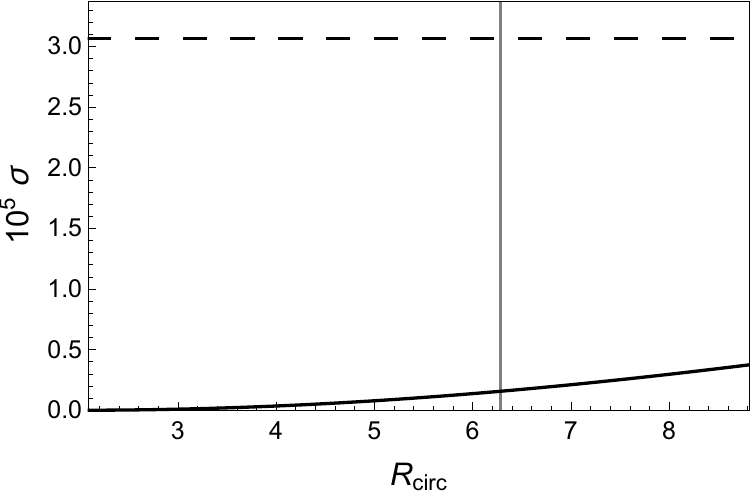}\quad
\includegraphics[width=0.48\columnwidth]{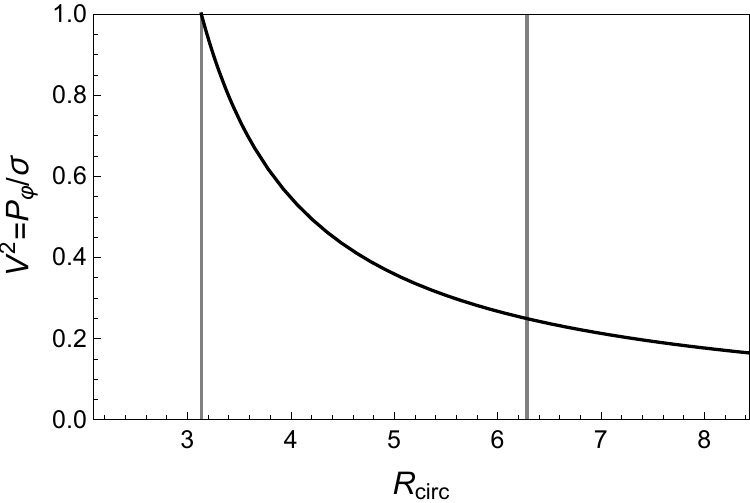}
\caption{
Same as Figure~\ref{fig:figN2a}, but for a $N=3$ ``Schwarzschild BH + disc'' system with parameters $m_1=1$, $m_2=-0.14$, $m_3=10$, $L_1=2$, $L_2=1$, $L_3=20$, $Z_1=\lambda=200$, $Z_2=176.35$, $Z_3=0$. We have $\sigma_{ISCO}/\sigma_{\rm max} \approx 0.051$. Compare with Figure~\ref{fig:figN3a}, whose configuration is the same but without the negative-mass rod.
}
\label{fig:figN6a}
\end{center}
\end{figure*}

\begin{figure*}
\begin{center}
\includegraphics[width=0.48\columnwidth]{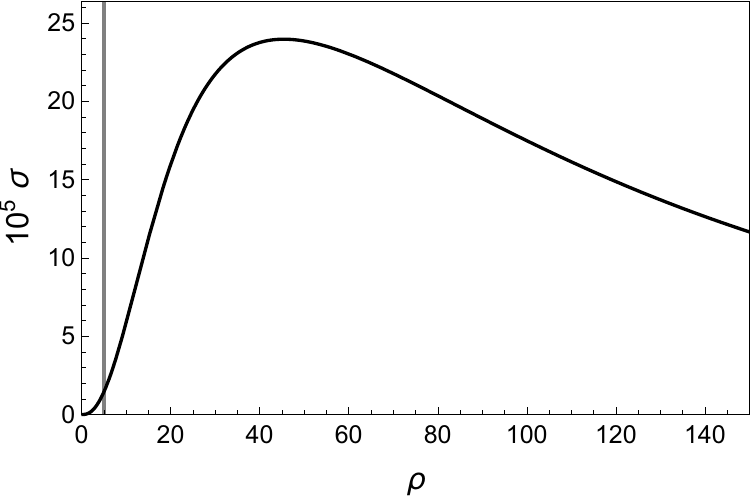}\quad
\includegraphics[width=0.48\columnwidth]{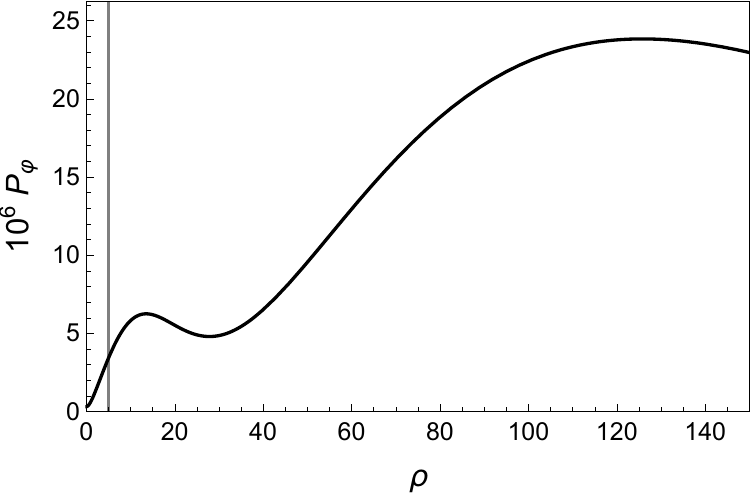}
\\ \hfill\\
\includegraphics[width=0.48\columnwidth]{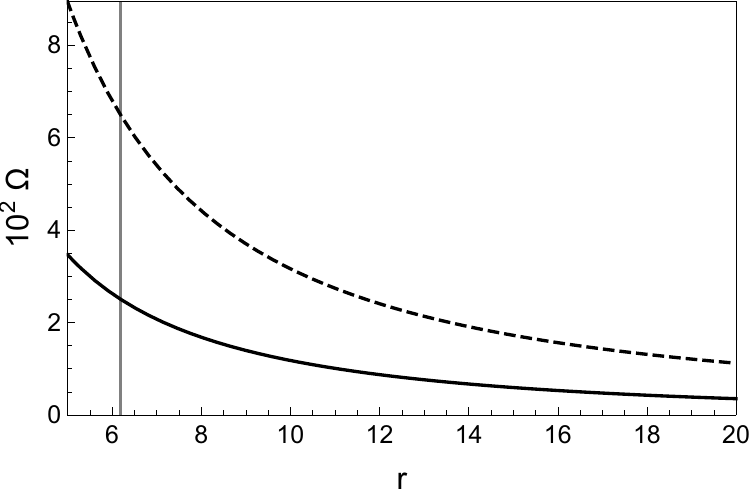}\quad
\includegraphics[width=0.48\columnwidth]{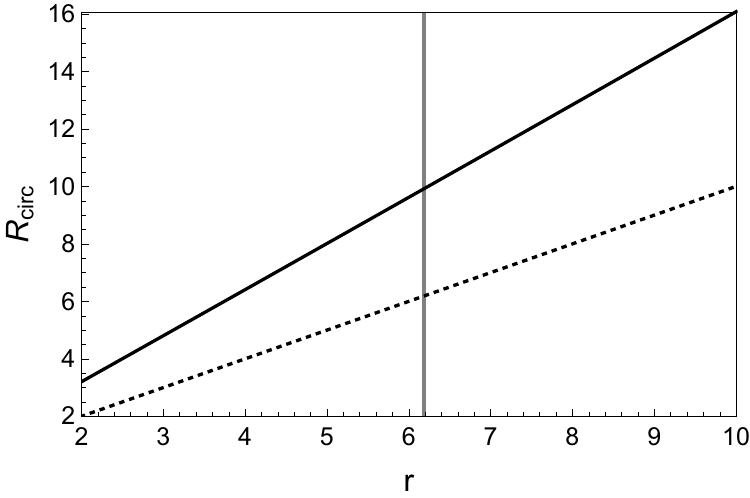}
\\ \hfill\\
\includegraphics[width=0.48\columnwidth]{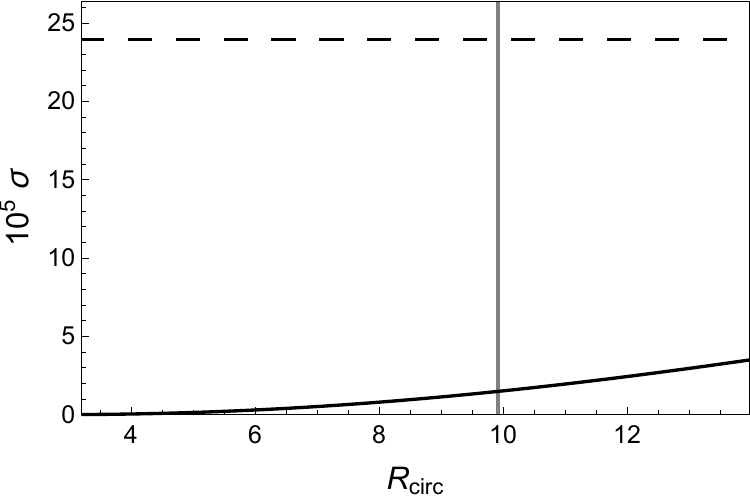}\quad
\includegraphics[width=0.48\columnwidth]{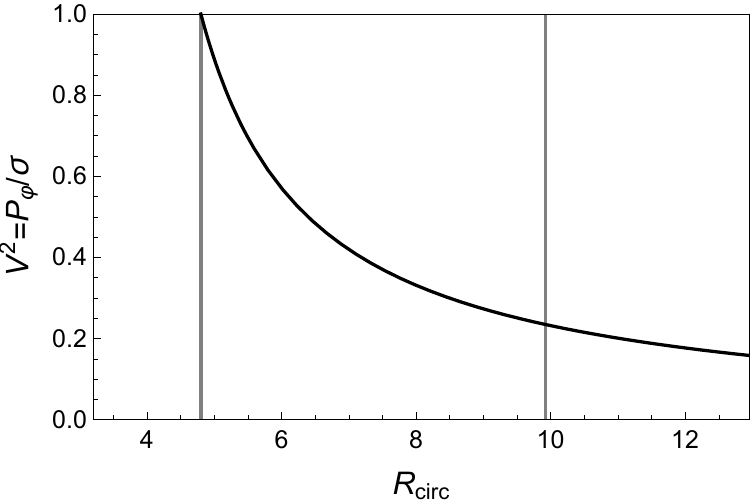}
\caption{
Same as Figure~\ref{fig:figN2a}, but for a $N=3$ ``Schwarzschild BH + disc'' system with parameters $m_1=1$, $m_2=-1.86$, $m_3=100$, $L_1=2$, $L_2=1$, $L_3=200$, $Z_1=\lambda=200$, $Z_2=176.35$, $Z_3=0$. We have $\sigma_{ISCO}/\sigma_{\rm max} \approx 0.06$. Compare with Figure~\ref{fig:figN4a}, whose configuration is the same but without the negative-mass rod.
}
\label{fig:figN7a}
\end{center}
\end{figure*}

In order to have a physically plausible disc, we must guarantee that $\sigma$ is negligible (in some sense) for $\rho<\rho_{ISCO}$ and that $P_\varphi(0)\approx 0$. However, the $N=2$ discs have, in general, a non-negligible surface density for $\rho<\rho_{ISCO}$ and an azimuthal pressure peaked at $\rho = 0$ (see Figures~\ref{fig:figN2a}, \ref{fig:figN3a}, and \ref{fig:figN4a}). We overcome this issue by considering at least a third rod in the seed solution, and this rod must have a negative mass parameter in order to diminish the disc density near its centre, while keeping $\sigma$ always non-negative (the ordering of the rods will depend on the situation; the only requisite is that at least one of the mass parameters must be negative). By choosing adequate parameters for the negative-mass rod, we can make the disc's surface density negligible for $\rho<\rho_{ISCO}$ and, at the same time, reduce the azimuthal pressure $P_\varphi(0)$ at the disc centre to arbitrarily low values. 
We show examples of this construction in Figures~\ref{fig:figN5a}, \ref{fig:figN6a}, and \ref{fig:figN7a} (we consider $m_2<0$  and $m_3>0$ for practical purposes, always with $L_2>0$, $L_3>0$). 
These discs were constructed by inserting an additional, negative-mass rod to the discs of Figures~\ref{fig:figN2a}, \ref{fig:figN3a}, and \ref{fig:figN4a}, respectively, so that we can compare them and estimate the rod's influence in their physical properties, particularly in the region near the BH.

It must be noted that, since the disc is composed of counterrotating, geodesic streams, the azimuthal pressure is generated by the circular motion of the disc particles. It follows that the relevant quantity which measures how negligible the disc is in the region $\rho<\rho_{ISCO}$ is indeed the surface density.
A practical approach, which we will adopt here, is to consider only models with sufficiently small surface density $\sigma_{ISCO} = \sigma(\rho_{ISCO})$ when compared to the maximum value of the surface density, $\sigma_{\rm max}$. For calculation purposes, we choose a threshold level $\sigma_{ISCO}/\sigma_{\rm max} < 10\%$ (we remark that this condition is coordinate independent). 
The above criterion was adopted to construct the discs of Figures~\ref{fig:figN5a}, \ref{fig:figN6a}, and \ref{fig:figN7a}. We will come back to the analysis of the region near the horizon at the end of this Section.

\begin{figure*}
\begin{center}
\includegraphics[width=0.48\columnwidth]{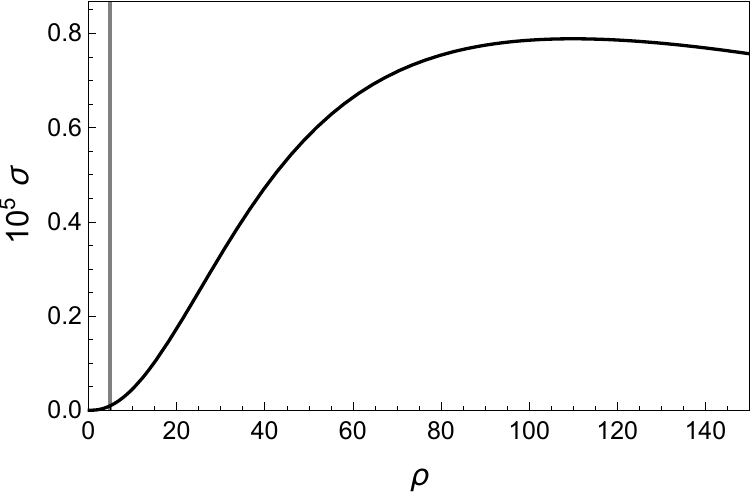}\quad
\includegraphics[width=0.48\columnwidth]{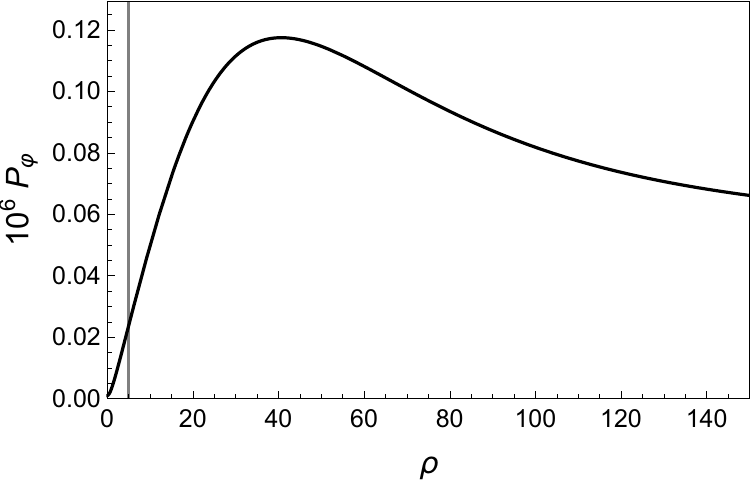}
\\ \hfill\\
\includegraphics[width=0.48\columnwidth]{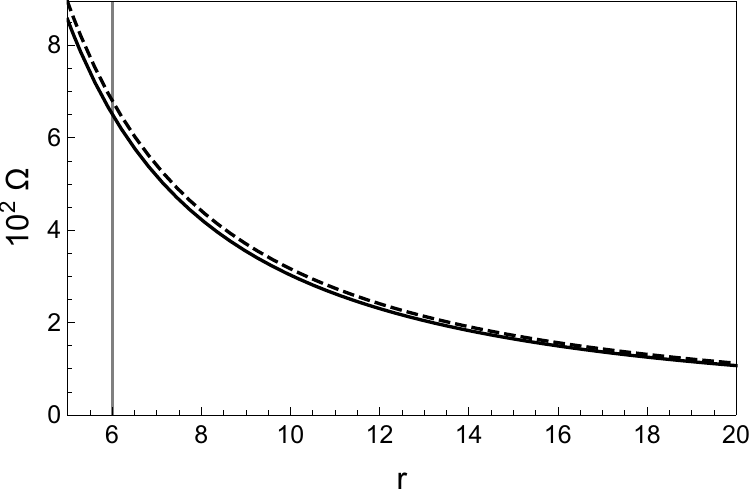}\quad
\includegraphics[width=0.48\columnwidth]{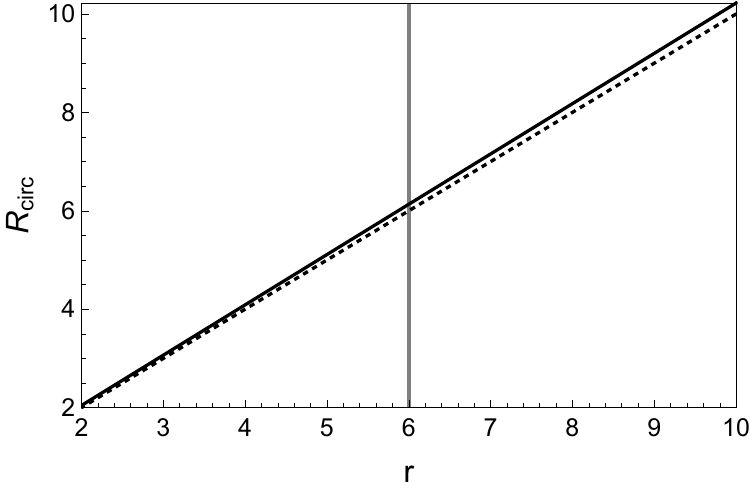}
\\ \hfill\\
\includegraphics[width=0.48\columnwidth]{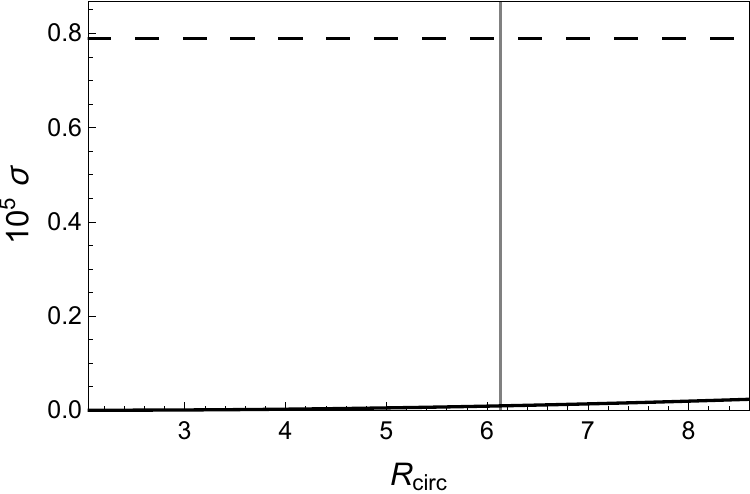}\quad
\includegraphics[width=0.48\columnwidth]{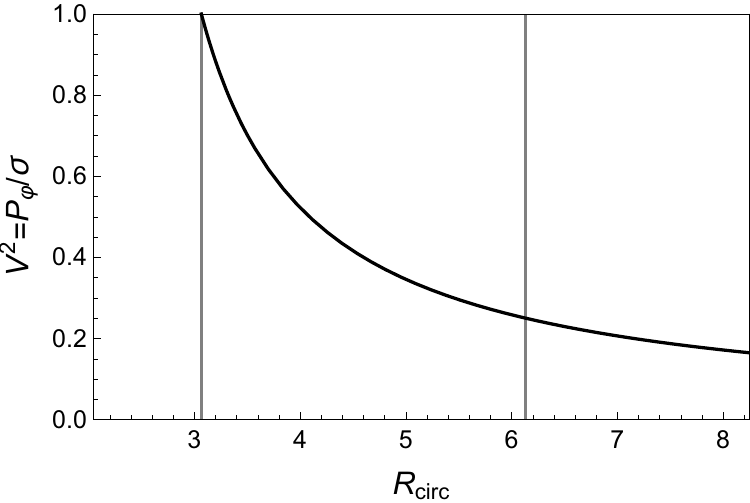}
\caption{
Same as Figure~\ref{fig:figN2a}, but for a $N=3$ ``Schwarzschild BH + disc'' system with parameters $m_1=1$, $m_2=-0.1563$, $m_3=10$, $L_1=2$, $L_2=1$, $L_3=20$, $Z_1=\lambda=400$, $Z_2=350$, $Z_3=0$. We have $\sigma_{ISCO}/\sigma_{\rm max} \approx 0.01$. We may see the difference in the surface density and pressure profiles compared to Figure~\ref{fig:figN6a}, for which $m_3 = 10$ but $Z_1=\lambda=200$.
}
\label{fig:figN8a}
\end{center}
\end{figure*}

Regarding the dependence of $\sigma$ and $P_\varphi$ on the parameters of the models, we may already infer some qualitative relations, even for the $N=2$ discs (top panels of Figures~\ref{fig:figN2a}, \ref{fig:figN3a}, and \ref{fig:figN4a}). We have two free parameters: $m_2$ and $Z_1-Z_2$. The effect of increasing the mass parameter $m_2$ is only to scale the surface density and azimuthal pressure profiles; the higher $m_2$, the higher the maximum of these profiles, keeping their general radial dependence (apart from a small increase of the azimuthal pressure for large radii when the mass parameter $m_2$ becomes very large with respect to $m_1$, as we see in Figure~\ref{fig:figN4a}). 
On the other hand, the higher the parameter $Z_1-Z_2$ for the $N=2$ discs (or $Z_1-Z_3$ for the $N=3$ discs), the higher the radius of maximum surface density and the lower the peak density, as exemplified in the top panels of Figure~\ref{fig:figN8a} with $Z_1-Z_3=400$ (compare with the corresponding panels of Figure~\ref{fig:figN6a}, which has the same positive-mass parameter $m_3 = 10$ but $Z_1-Z_3=200$). Since we are dealing essentially with an image method, this fact may be understood by noting that if the image source is far from the surface where the disc will be created, the corresponding gravitational field will be weaker.
Moreover, the coordinate radius of maximum surface density is always larger in the $N=3$ discs when compared to the corresponding $N=2$ discs.

\subsection{Effects of the disc's self gravity}\label{sec:selfgravity}

Let us now consider the Schwarzschild-like coordinates ($r,\theta$) given by $\rho=\sqrt{r(r-2m_1)\,}\sin\theta$, $z=(r-m_1)\cos\theta$ \cite{semerakZellerinZacek1999MNRAS, saaVenegeroles1999PhLA, griffithsPodolsky2009exact}; the event horizon is located at the hypersurface $r = 2 m_1$.
As a first way of quantifying the influence of the disc's self gravity on test-particle motion, we plot in Schwarzschild-like coordinates the angular speed profile of test particles $\Omega(r) = \sqrt{-\partial_r g_{tt}/\partial_r g_{\varphi\varphi}\,}$ on the equatorial plane. We compare it with the Keplerian expression (obtained in the absence of the disc), $\Omega_K(r) = \sqrt{m_1/r^3\,}$. We have that, asymptotically, $\Omega(r)\sim \sqrt{M_{\rm ADM}/r^3\,}$ as expected, implying $\Omega>\Omega_K$ for large radii (since we only consider cases in which $M_{\rm ADM}>m_1$) with the same radial dependence. However, the profile for radii comparable or smaller than the radius of maximum surface density differs considerably from the Keplerian curve, as we see in the middle, left panel of Figures~\ref{fig:figN2a}, \ref{fig:figN3a}, \ref{fig:figN4a}, \ref{fig:figN5a}, \ref{fig:figN6a}, and \ref{fig:figN7a}. For these radii we have $\Omega<\Omega_K$, with a larger difference as we approach the hypersurface $r=2m_1$. This difference increases with increasing $M_{\rm ADM}$; systems with $M_{\rm ADM} \gg m_1$ may give us a considerable difference between the $\Omega(r)$ and $\Omega_K(r)$ profiles (see for instance Figures~\ref{fig:figN4a} and \ref{fig:figN7a}). This effect is only slightly affected by diminishing $\sigma$ and $P_\varphi$ near the disc centre, as we see by comparing the middle, left panels of the mentioned Figures for the same positive-mass parameter of the disc ($m_2$ for $N=2$ and $m_3$ for $N=3$; compare Figures~\ref{fig:figN2a} and \ref{fig:figN5a}, \ref{fig:figN3a} and \ref{fig:figN6a}, \ref{fig:figN4a} and \ref{fig:figN7a}). The effect of the negative-mss rod is more prominent for large $m_3$ (as exemplified in Figures~\ref{fig:figN4a} and \ref{fig:figN7a}). 

Also, we know that, neglecting self gravity,  deviations from the Keplerian angular velocity appear only for accretion discs with high accretion rates; in this case the thin discs thicken and become slim \cite{abramowiczEtal1988ApJ, sadowskiEtal2011AA, lasotaVieiraEtal2016AA}. But, as shown here, self gravity may introduce deviations from the Keplerian profile even in the thin-disc regime, having an impact on the physical properties of the corresponding accretion disc models. This analysis is, however, beyond the scope of the present work.

\begin{figure*}
\begin{center}
\includegraphics[width=0.48\columnwidth]{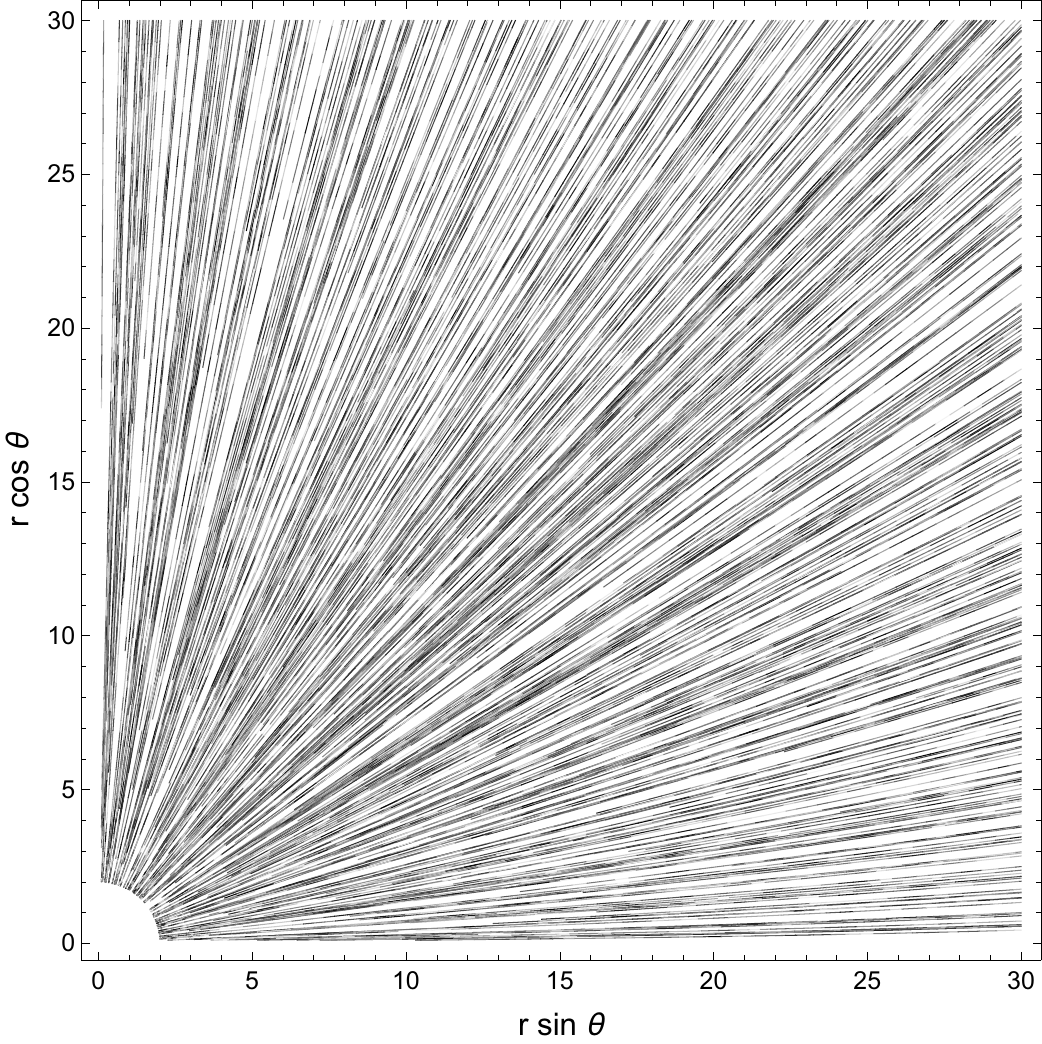}\quad
\includegraphics[width=0.48\columnwidth]{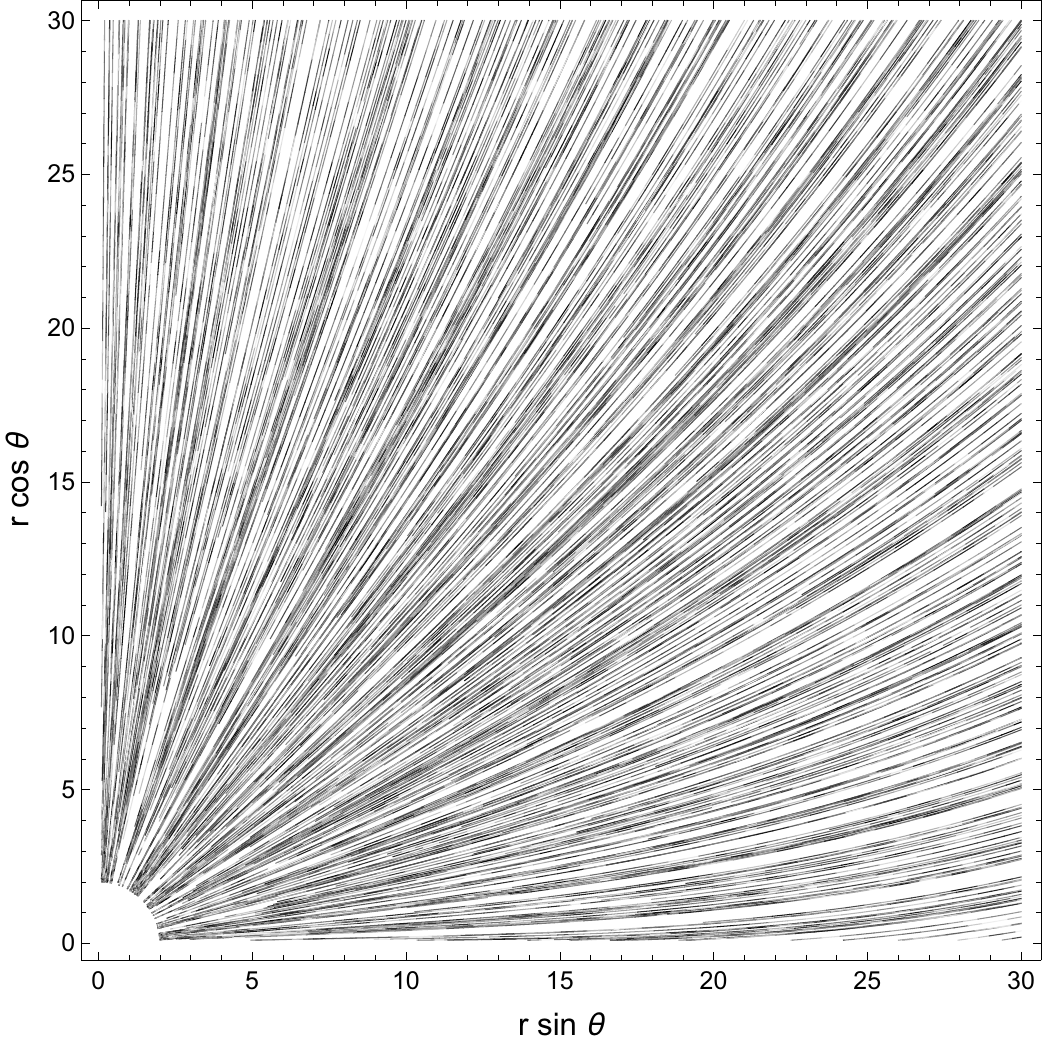}\\ \hfill \\
\includegraphics[width=0.48\columnwidth]{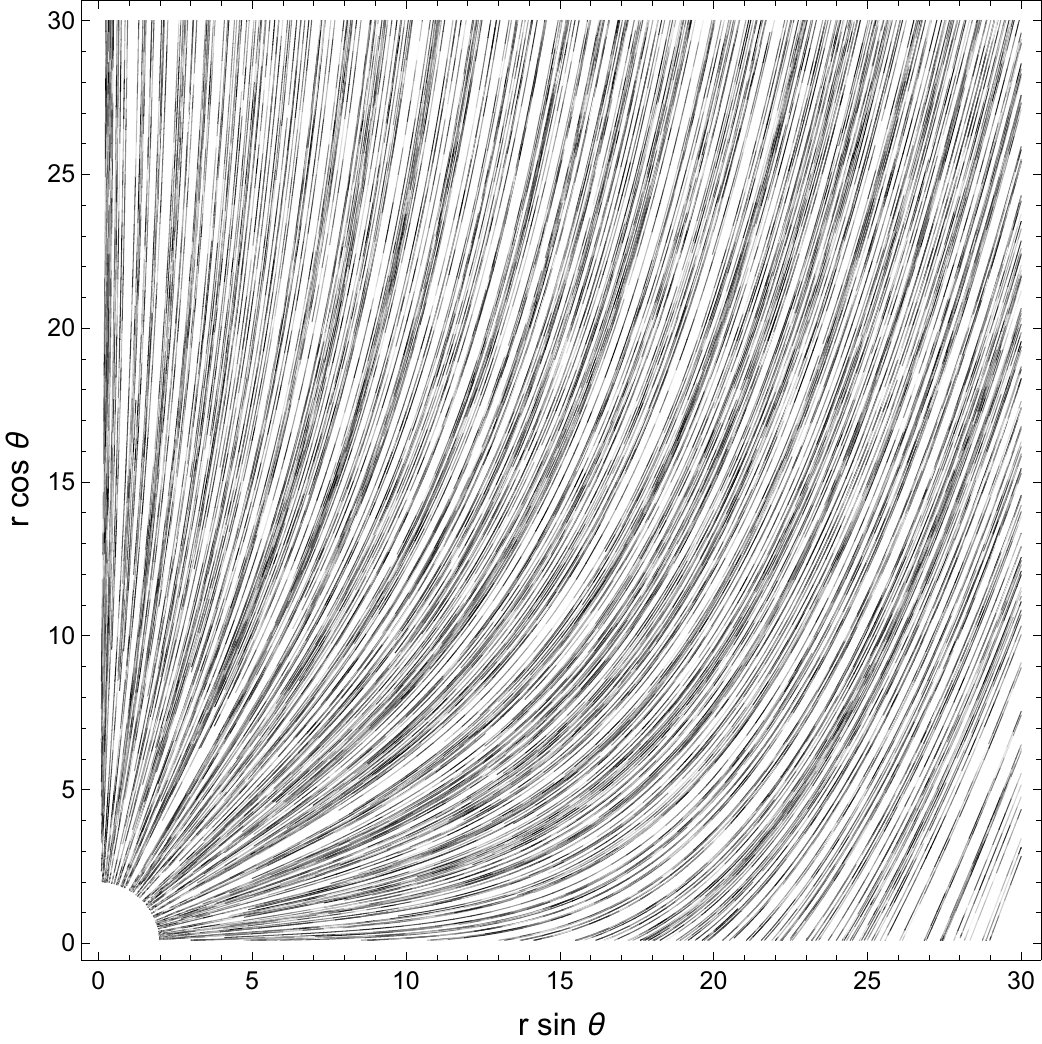}\quad
\includegraphics[width=0.48\columnwidth]{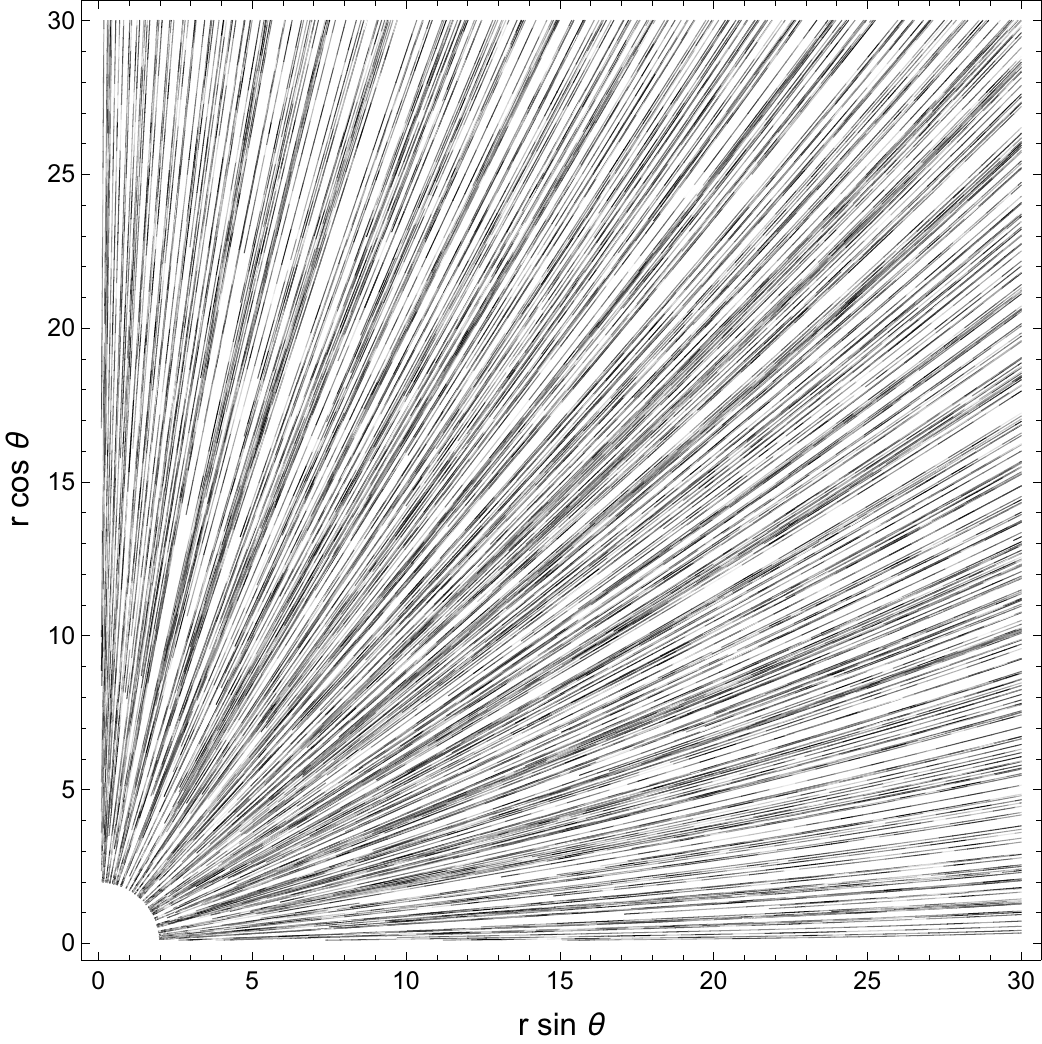}\quad

\caption{
Field lines of the $N=3$ discs in Schwarzschild-like coordinates ($r,\theta$). Top, left: disc of Figure~\ref{fig:figN5a}. Top, right: disc of Figure~\ref{fig:figN6a}. Bottom, left: disc of Figure~\ref{fig:figN7a}. Bottom, right: disc of Figure~\ref{fig:figN8a}. The distortion of the field lines (and consequently the disc influence on off-equatorial particle motion) increases with increasing $m_3$, as seen in the first three panels, and decreases with increasing $Z_1-Z_3$, as shown in the last panel.
}
\label{fig:figN9a}
\end{center}
\end{figure*}

We also analyze the system's field lines (the stream lines of the 4-acceleration $a_\mu=u^\nu\nabla_\nu u_\mu$ of a static observer) in Schwarzschild-like coordinates ($r,\theta$). It can be shown that, for a static observer, $a_\mu=\partial_\mu\psi$ \cite{semerakZellerinZacek1999MNRAS}, and therefore the field lines are perpendicular to the contour lines of $\psi$. In the absence of the disc, the field lines present spherical symmetry. The distortion of the field lines from spherical symmetry is then a qualitative view of the disc influence on test-particle dynamics; field lines extending up to the equatorial plane represent the dominant contribution of the disc to the 4-acceleration of static observers, as we exemplify in Figure~\ref{fig:figN9a} for the $N=3$ discs of Figures~\ref{fig:figN5a}, \ref{fig:figN6a}, \ref{fig:figN7a}, and \ref{fig:figN8a}. We see that the disc influence on off-equatorial particle motion increases with increasing $m_3$ (the positive-mass parameter), and decreases with increasing $Z_1-Z_3$; for the $m_3 = 1$ disc (Figure~\ref{fig:figN5a}), the distortion of the field lines is negligible, as it also happens for the $Z_1-Z_3 = 400$ disc of Figure~\ref{fig:figN8a} (see the bottom, right panel of Figure~\ref{fig:figN9a}). Comparing with the behaviour of the surface density, we may state that the disc influence on off-equatorial motion increases with increasing surface density, as expected from the third-integral formalism of \cite{vieiraRamosCaro2016CeMDA, vieiraRamoscaroSaa2016PRD}.

\subsection{The region near the horizon}\label{sec:nearHorizon}

Before looking at the behaviour of the physical quantities of the disc near the horizon, we ask whether coordinate effects affect significantly our discussion. In order to do this, we must consider the behaviour of the circumferential radius in terms of the Schwarzschild-like coordinates ($r,\theta$). 
The circumferential radius on the equatorial plane $z=0$ ($\theta = \pi/2$) is defined as $R_{\rm circ} = \sqrt{g_{\varphi\varphi}\,}$ \cite{thorne1977ApJ, kotlarikSemerakCizek2018PRD, katkaVieiraEtal2015GRG}, in such a way that $2\pi\,R_{\rm circ}$ is the corresponding proper circumference. In Schwarzschild spacetime, it coincides with the usual coordinate `$r$' on the equatorial plane. Moreover, $g_{\varphi\varphi}$ has a geometric meaning given by the squared norm of the spacelike Killing vector field $\xi = \partial/\partial\varphi$, that is $g_{\varphi\varphi} = \xi^2 = g_{\mu\nu}\,\xi^\mu \xi^\nu$ \cite{abramowiczKluzniak2005ApSS}.  
So the circumferential radius is the more reliable way to compare quantities in distorted BH fields with quantities in Schwarzschild spacetime. It is indeed widely used as a geometrical measure of the shape of distorted BH horizons \cite{semerakZellerinZacek1999MNRAS, semerakZacek2000CQGra} and has applications in the physics of neutron stars \cite{thorne1977ApJ, thorneZytkow1977ApJ}. 
In particular, the location of the ISCO in distorted BH spacetimes only makes sense if we refer to its circumferential radius $R_{ISCO}$, and not to the coordinate radius $\rho_{ISCO}$. Then we may compare the geometric ISCO radii of the `pure' and distorted BH spacetimes and find whether the ISCO radius shrinks or expands in the presence of the surrounding gravitating disc. 

We show in the middle, right panel of Figures~\ref{fig:figN2a}, \ref{fig:figN3a}, \ref{fig:figN4a}, \ref{fig:figN5a}, \ref{fig:figN6a}, and \ref{fig:figN7a} the dependence of the circumferential radius on the Schwarzschild-like coordinate radius $r = m_1 + \sqrt{m_1^2 + \rho^2\,}$ (expression valid on the equatorial plane). In the absence of the disc, we have $R_{\rm circ} = r$\,. 
The general picture is that $R_{\rm circ}> 2 m_1$ at the horizon and grows roughly linearly as a funtion of $r$, but with an average angular coefficient greater than unity. The average angular coefficient increases with increasing $M_{\rm ADM}$. The difference in the behaviour of $R_{\rm circ}(r)$ between the $N=2$ and $N=3$ discs (with the same positive-mass parameter) is negligible, with small differences for large masses $m_2$ ($N=2$) and $m_3$ ($N=3$), as we see by visually comparing the corresponding panels of Figures~\ref{fig:figN4a} and \ref{fig:figN7a}. Moreover, for growing values of $Z_1-Z_2$ (for $N=2$) or $Z_1-Z_3$ (for $N=3$) we have that the average angular coefficient of $R_{\rm circ}(r)$ decreases, regardless of the negative-mass rod, approaching unity when the parameter gets very large (as exemplified in Figure~\ref{fig:figN8a}). 
In particular, we find that for all the discs obtained $R_{ISCO}> 6m_1$, showing that the circumferential ISCO radius increases in the presence of the gravitating disc, although the corresponding Schwarzschild-like radial coordinate $r_{ISCO}$ may be larger or smaller than $6m_1$, depending on the disc parameters. We remark that this phenomenon happens not only in the $N=2$ case but also in the $N=3$ case, when the surface density at the ISCO radius is made much smaller.

Coming back to the radial dependence of $\sigma$ near the horizon, we plot the surface density against circumferential radius in the bottom, left panel of Figures~\ref{fig:figN2a}, \ref{fig:figN3a}, \ref{fig:figN4a}, \ref{fig:figN5a}, \ref{fig:figN6a}, and \ref{fig:figN7a}, focusing on that region. For the $N=2$ discs, as mentioned before, $\sigma_{ISCO}/\sigma_{\rm max}$ is large. 
The parameters for the $N=3$ discs presented here were obtained by fixing $m_1, L_1, Z_1$ and $m_3, L_3, Z_3$, and then varying the parameters of the negative-mass rod $m_2, L_2, Z_2$ until the inequality $\sigma_{ISCO}/\sigma_{\rm max} < 10\%$ is obtained (see Figures~\ref{fig:figN5a}, \ref{fig:figN6a}, and \ref{fig:figN7a}), and also until diminishing the central azimuthal pressure to low values.
We remark that there is no analytical criterion to find the parameters which guarantee the condition $\sigma_{ISCO}/\sigma_{\rm max} < 10\%$. 

The subtle question of superluminal speeds for the counterrotating disc matter near the central BH deserves a deeper discussion. When we construct a `BH + disc' spacetime from a $N$-rod seed model as done here, we will have in general $P_\varphi(0)>0$ and $\sigma(0)=0$. As mentioned above for the $N=3$ family, a small negative mass $m_3<0$ may also diminish the value of the pressure, but since the analytical expression for  $P_\varphi$ is too cumbersome even for the $N=2$ case, there is no algorithmic method to make $P_\varphi\to 0$ as $\rho\to 0$. Also, even if one obtains this condition, it is not clear whether we could have $P_\varphi/\sigma$ bounded when approaching the horizon.
This limitation of the method will then produce, in general, discs with superluminal counterrotating speeds near the BH (it is indeed a general limitation of the DCR method, which forces a discontinuity in the whole equatorial plane, and not specifically of this seed solution). 

We evaluate the circumferential radius where the counterrotating streams have $V^2=1$, which gives us the photon orbit locus of the spacetime. We plot $V^2 = P_\varphi/\sigma$ as a function of $R_{\rm circ}$ for each disc in the bottom, right panel of Figures~\ref{fig:figN2a}, \ref{fig:figN3a}, \ref{fig:figN4a}, \ref{fig:figN5a}, \ref{fig:figN6a}, and \ref{fig:figN7a}. It is clear that the circumferential radius corresponding to $V^2=1$ is always larger than $3m_1$, the corresponding value in the absence of the disc, as it happens with the ISCO radius $R_{ISCO}$ (when compared to $6m_1$). 
Moreover, since the $V^2=1$ circumferential radius is always smaller than $R_{ISCO}$, and accordingly the surface density $\sigma_1$ at the $V^2=1$ radius satisfies $\sigma_1<\sigma_{ISCO}$, there would be a very small amount of particles near the BH with the undesired property mentioned above.  More elaborate models, considering other seed sources of the form ``BH + source'' and with the DCR transformation applied the same way as here (see Figure~\ref{fig:DCRmethod}), may circumvent this imperfection, if one is able to eliminate the discontinuity of the metric generated by the DCR method near the horizon. This is, however, beyond the scope of the present work.

All the above arguments lead us to the conclusion that we may neglect the disc contribution for $\rho<\rho_{ISCO}$ in the $N=3$ case by a suitable choice of parameters, considering then an ``effective'' inner rim for the disc close to $\rho_{ISCO}$. We then justify its annular character.

\section{Charged discs around extremal BHs}
\label{sec:Charged}

The formalism presented here can be readily applied to the Majumdar-Papapetrou-type seed solution of $N$ collinear extremal R-N BHs \cite{majumdar1947PR, griffithsPodolsky2009exact}, each one at position $Z_i$ along the $z$-axis with $Z_1>...>Z_N$. After the DCR transformation (\ref{eq:DCRtransformation}) with $\lambda=Z_1>0$, the metric becomes
\begin{equation}\label{eq:ConformastaticMetric}
ds^2=-U^{-2}\,dt^2 + U^2\,(d\rho^2 + \rho^2 d\varphi^2 + dz^2)\, ,
\end{equation}
where
\begin{equation}\label{eq:PotentialNChargedParticles}
U=1+ \frac{M_1}{\sqrt{\rho^2 + z^2}} + \sum_{i=2}^N\left(\frac{M_i}{\sqrt{\rho^2 + \zeta_i^2}}\right)
\end{equation}
and $\zeta_i= |z| + Z_1 -Z_i$. The corresponding 4-potential is $A_\mu=(1/U,0,0,0)$ \cite{griffithsPodolsky2009exact, ryznerJofka2015CQG}.
The system represents an extremally charged dust disc ($\sigma_e=\sigma$, see \cite{loraclavijo-ospinahenao-pedraza2010PRD}) around an extremal R-N BH of mass $M_1$. It must be noted that, comparing metric (\ref{eq:ConformastaticMetric}) with the general Weyl metric (\ref{eq:Weylmetric}), we have $\gamma=1$ identically and therefore $P_\varphi = 0$ (see Equation~(\ref{eq:PphiWeyl})).
The disc's surface density profile (\ref{eq:sigmaWeyl}) has a simple expression given by 
\begin{equation}\label{eq:sigmaChargedNsources}
\sigma = \frac{\sum_{i=2}^N\frac{M_i (Z_1 -Z_i)}{\left((Z_1 -Z_i)^2+\rho^2\right)^{3/2}}}{2 \pi  \left(1+\frac{M_1}{\rho}+\sum_{i=2}^N\frac{M_i}{\sqrt{(Z_1 -Z_i)^2+\rho^2}}\right)^2 }\,.
\end{equation}
It vanishes at the origin of Weyl coordinates (corresponding to the event horizon), $\sigma(0)=0$, and therefore the disc also has annular character; it also has a $\sigma\sim 1/\rho^3$ power-law tail.

We mention that the DCR method can generate a strut-free equilibrium configuration of two equal-mass extremal R-N BHs  with a charged razor-thin disc in its mid-plane, by taking $Z_2<\lambda<Z_1$ in the DCR method. The BHs will both have mass $M_1$ and will be located at $z=\pm (Z_1-\lambda)$. If we take instead $\lambda>Z_1$ we will have a disc-only solution. The corresponding expressions for the DCR-transformed metric and for the disc surface density follow directly from the formalism.
In both cases the discs are infinite in extent and satisfy $\sigma>0$ for $\rho>0$.

\section{Discussion}
\label{sec:Discussion}

We presented in this paper exact solutions of Einstein's field equations corresponding to Schwarzschild BHs surrounded by annular, counterrotating razor-thin discs of infinite extent.
The superposed ``Schwarzschild BH + disc'' solutions were obtained via the application of the DCR method (Section \ref{sec:RTDs}) to seed solutions representing $N$ collinear Weyl rods \cite{letelierOliveira1998CQG}, the extension of the corresponding $N$-BH solution \cite{israelKhan1964NCim}.
We remark that, by considering $L_1\neq 2m_1$ in the DCR procedure with $\lambda=Z_1$ (Section \ref{sec:AnnularSchwarzschild}), we obtain annular discs around an arbitrary Weyl rod. 


Although we considered only the $N=3$ case for presentation purposes, the procedure of constructing razor-thin discs around Schwarzschild BHs via multi-hole seeds allows for an arbitrary (but finite) number of Weyl rods in the seed spacetime, generating an infinite family of razor-thin discs around Schwarzschild BHs. The formalism can also be extended to more general seed spacetimes constituted of a BH plus an arbitrary axially symmetric source, such as for instance an additional Schwarzschild BH supported by an Appell ring \cite{semerakEtal2019PRD} or a line source of variable density \cite{bicakLyndenbellKatz1993PRD} below the Schwarzschild BH, possibly avoiding the superluminal counterrotating matter near the BH.
The crucial step of the method, which gives us a central Schwarzschild BH, is to choose the DCR parameter $\lambda$ in such a way that the corresponding hypersurface `cuts in half' the highest Schwarzschild rod (in Weyl coordinates), as depicted in Figure~\ref{fig:DCRmethod}.  


The stability analysis of the disc presented here takes into account only the linear stability of the corresponding circular geodesics, once we interpret these discs as being composed of counterrotating particles \cite{lemosLetelier1994PRD, gonzalezEspitia2003PRD}. The discs are vertically \cite{vieiraRamoscaroSaa2016PRD} and radially \cite{letelier2003PRD} stable according to this approximation. However, the disc fluid elements may not follow geodesics, since the disc has azimuthal pressure; this pressure must be taken into account in the stability analysis of the fluid \cite{freitasSaa2017PRD}. A complete treatment would involve the linearised conservation laws for small perturbations to the disc fluid quantities (density and pressure) such as in \cite{ujevicLetelier2004PRD, ujevicLetelier2007GRG, ujevicLetelier2007MNRAS}. This kind of analysis is beyond the scope of the present work.

We also obtained extremally charged dust discs around extremal R-N BHs as another application of the method.  
We may then extend our results to (neutral or charged) discs around R-N BHs with arbitrary charge-to-mass ratio, following for instance the seed solution of \cite{azumaKoikawa1994PThPh}. 
Another possibility is to construct stationary discs around Kerr BHs via the multi-hole DCR method presented here, for instance from the double-Kerr solution \cite{kramerNegembauer1980PLA, letelierOliveira1998PLA}. These issues are left for future work.

%

%
\section*{Acknowledgments}
R.S.S.V. thanks Alberto Saa for stimulating discussions. The author acknowledges the anonymous referees for the many helpful suggestions which allowed to improve the original manuscript.
This work was partially supported by the Coordena\c{c}\~ao de Aperfei\c{c}oamento de Pessoal de N\'ivel Superior - Brasil (CAPES) - Finance Code 001.
%


\newcommand{\newblock}{}

\end{document}